\documentclass[12pt, letterpaper]{article}

\usepackage{amsmath}
\usepackage{amsfonts}
\usepackage{amssymb}
\usepackage{graphicx}

\usepackage{color}

\usepackage{amsmath, amssymb, graphics, epsfig, graphicx, epstopdf}

\usepackage{amsfonts}
\usepackage{amssymb}
\usepackage{graphicx}
\usepackage{amstext}
\usepackage{color}

\newcommand{\nc}{\newcommand}
\nc{\ba}{\begin{eqnarray}}
\nc{\ea}{\end{eqnarray}}

\newcommand{\bea}{\begin{eqnarray}}
\newcommand{\eea}{\end{eqnarray}}

\nc{\be}{\begin{eqnarray}}
\nc{\ee}{\end{eqnarray}}

\nc{\bfk}{{\bf k }}
\nc{\bfx}{{\bf x }}
\nc{\pfp}{{\bf{p}}}
\nc{\bfp}{{\bf{p}}}
\nc{\bfq}{{\bf{q}}}
\nc{\tbf}{\textbf}

\nc{\calP}{  { \cal P} }
\nc{\calR}{  { \cal R} }
\nc{\im}{ \mathrm{Im} }
\nc{\sg}{ \mathrm{sgn} }
\begin{document}

\vspace{5mm}
\vspace{0.5cm}
\begin{center}

\def\thefootnote{\fnsymbol{footnote}}

{ \large {\bf Effective field theory of statistical anisotropies\\ for  primordial bispectrum and gravitational waves}   }
\\[0.5cm]

{ Tahereh Rostami$^{1}$\footnote{t.rostami@ipm.ir}, Asieh Karami$^{1}$\footnote{karami@ipm.ir}, Hassan Firouzjahi$^{1}$\footnote{firouz@ipm.ir }}
\\[0.5cm]

{\small \textit{$^1$School of Astronomy, Institute for Research in Fundamental Sciences (IPM) \\ P.~O.~Box 19395-5531, Tehran, Iran
}}\\

\end{center}

\vspace{.8cm}

\hrule \vspace{0.3cm}


\begin{abstract}

We present the effective field theory studies of primordial statistical anisotropies in models of anisotropic inflation. The general action in unitary gauge is presented  to calculate the 
leading interactions between the gauge field fluctuations, the curvature perturbations and the tensor perturbations.  The anisotropies in scalar power spectrum and bispectrum are calculated and the  dependence of these anisotropies  to EFT couplings are presented. In addition, we calculate the statistical anisotropy in tensor power spectrum and the scalar-tensor  cross correlation. Our EFT approach 
incorporates anisotropies generated in models  with  non-trivial speed for the gauge field fluctuations and sound speed for scalar perturbations such as  in DBI inflation.

\end{abstract}

\vspace{0.5cm} \hrule
\def\thefootnote{\arabic{footnote}}
\setcounter{footnote}{0}
\newpage
\section{Introduction}

There are many scenarios of inflation which are compatible with cosmological observations. With more precise observations to come in future one may hope to discriminate various inflationary scenarios and hopefully 
narrow down the landscape of observably viable scenarios of inflation. Having said this, it seems unrealistic if one can single out a particular model as the true realization of inflation in early universe.
This naturally raises the question if one can classify the various inflationary scenarios either based on their main predictions or based on their theoretical constructions.

The method of effective field theory (EFT) of inflation \cite{Cheung:2007st, Cheung:2007sv} was a successful program to answer some of the above questions in classifying various inflationary scenarios based on their predictions for power spectrum and bispectrum. From the 
view point of  EFT all interactions which are  compatible with the underlying symmetries should be considered (for a general review of EFT, see \cite{Manohar:1996cq, Burgess:2007pt}).  Then the predictions of  different inflationary models are realized  depending on which  interactions governing the dynamics of  the light fields are turned on. In particular the EFT approach was more  successful in dealing with single field models of inflation. This is mainly because in single field scenarios the inflaton field $\phi(t)$ plays the role of time so upon going to comoving gauge, i.e. choosing 
the time slicing as the surface of constant $\phi$, one can eliminate the fluctuations of the inflaton field. Consequently, the remaining degrees of freedom are mainly geometrical  in nature. One then is able to classify all the relevant interactions based on the number of derivatives involved which are allowed by the 
remaining three-dimensional time dependent diffeomorphisms \cite{Cheung:2007st}.

The models of inflation are mainly based on scalar field dynamics. The  scalar fields have spin-zero by construction so they are quite apt to construct isotropic cosmological backgrounds  as required by  cosmological principles. Indeed, various  cosmological observations indicate the universe as a whole seems to be isotropic to very high accuracies \cite{Ade:2013nlj, Ade:2015hxq}. Having said this,  it is natural to examine the role of fields with other spins during inflation. More specifically vector fields and gauge fields appear in abundant in  Standard Model of
particle physics and in quantum field theory. It is natural to expect that they play some roles during inflation.  Of course, because of the near isotropy of the universe, the vector fields and gauge fields by themselves may not play the role of inflaton fields. However, it is conceivable that they play the role of isocurvature light fields which may also be coupled to inflaton field. This brings the interesting possibilities that light gauge fields may affect the cosmological observations by generating some observable amount of statistical anisotropies. Anisotropic inflation 
is such a realization based on dynamics of gauge fields during inflation.  In the  most well-studied  realization of anisotropic inflation, a $U(1)$ gauge field with a non-zero electric field energy density is present during inflation. As typical of the vector field dynamics, the electric field energy density is quickly redshifted in an expanding cosmological background.  Therefore, in order to sustain the background electric field energy density and also in order to generate a scale-invariant  power spectrum for the  gauge field perturbations,  the gauge field is conformally coupled to the inflaton field.   These models of anisotropic inflation usually predict a small amount of quadrupolar statistical anisotropies on CMB maps
which can be  tested observationally \cite{Kim:2013gka, Planck:2013jfk, Ade:2015lrj}. 

The motivation of this work is to study the general setup of anisotropic inflation using the approach of  EFT of inflation. A similar study was performed in \cite{Abolhasani:2015cve} in which the dominant interactions generating statistical anisotropies in power spectrum were classified. See also 
\cite{Cannone:2015rra, Hidaka:2014fra, Lin:2015cqa, Bartolo:2015qvr} in which the EFT of anisotropies in the setups where all or some spacetime diffeomorphisms are broken is studied. 
The method employed in \cite{Abolhasani:2015cve} was similar to the approach proposed in  \cite{Cheung:2007st} in which upon fixing  time  by surface of constant $\phi(t)$, the dominant interactions are constructed from  $\delta g^{00}$ and other leading geometrical operators. Then, taking  the gauge symmetry into account, the operators invariant under  the  remnant symmetries  are constructed. This  was a novel method which proved efficient when dealing with  quadratic action to calculate the  anisotropic power spectrum. However, the above geometric approach becomes somewhat inadequate when one considers cubic and higher order terms in the action. We postpone the discussion associated with this difficulty  to section \ref{old approach}. Here we take a somewhat more practical approach as follows. Following the logic of \cite{Cheung:2007st} we still fix the time coordinate by choosing the surface of constant $\phi(t)$. Then we simply write down the leading interactions allowed for the gauge field fluctuations. In a sense, this approach is a hybrid of the EFT in multiple field scenarios \cite{Senatore:2010wk} and the EFT approached employed by Weinberg \cite{Weinberg:2008hq}. This approach allows us to calculate the cubic interactions and  the anisotropic bispectrum. In addition, we also study the tensor perturbations and the scalar-tensor cross correlation generated in anisotropic background within our EFT approach.

\section{EFT of Anisotropic Inflation}
\label{eft-anisotropic inflation}
In this section we review the setup of anisotropic inflation and then present the general EFT action in unitary gauge.  For a review on anisotropic inflation see \cite{Emami:2015qjl} and for 
various works related to anisotropic  power spectrum and bispectrum and their observational imprints on CMB and large scale structure 
 see \cite{Watanabe:2009ct, Watanabe:2010fh, Soda1, Emami1, Emami:2013bk, Abolhasani:2013zya, Abolhasani:2013bpa,  Chen:2014eua,  Bartolo:2012sd, Dulaney:2010sq, Shiraishi:2013vja, Shiraishi:2013oqa, Barnaby:2012tk,   various, Shiraishi:2016wec, Naruko:2014bxa}. See also  \cite{various2, Emami:2016ldl, Mukohyama:2016npi, Fleury:2014qfa, Talebian-Ashkezari:2016llx}
for different realizations of statistical anisotropies.  

\subsection{Anisotropic inflation background }
\label{anisotropic inflation}

As mentioned before, in the setup of anisotropic inflation we have the scalar field $\phi$ as the  inflaton field and a $U(1)$ gauge field $A_\mu$ which is the source of electric field energy density during inflation. Without loss of generality we assume that the background electric field is along the $x$ direction  so the gauge field has the form $ A_\mu = (0,{A}_x(t),0,0)$. The background electric field energy density breaks the isotropy  so the background geometry is in the form of Bianchi type I  universe. However, the setup still has  the rotational symmetry in two-dimensional  $yz$ plane. 

Because of the conformal symmetry associated with the Maxwell theory, the background electric field energy density is diluted  if the gauge field is not coupled to inflaton.  Therefore, in order for the background electric field energy density to  survive  the dilution from  the exponential expansion,  the gauge field is coupled to the inflaton field. With the minimal extension of Maxwell theory, this coupling is given by   $-f(\phi)^2 F_{\mu \nu} F^{\mu \nu}/4$. The next goal is to choose the functional form of $f(\phi)$ such that the background electric field energy density to furnish a nearly constant but sub-leading fraction of the total energy density.  As shown in \cite{Watanabe:2009ct} this is an attractor system, in the sense that the system reaches the final stage in which the gauge field energy density furnishes a constant but sub-leading fraction of the inflaton field energy density.  For a given inflaton potential $V(\phi)$ the form of $f(\phi)$ which results in the above mentioned attractor phase 
can be obtained. In terms of scale factor $a(t)$, it takes the time-dependent value $f(\phi) \propto a(t)^{-2}$.   At the perturbation level, this choice of $f(\phi)$ also yields a  scale invariant power spectrum for the gauge field fluctuations. Calculating the interactions between inflaton and the gauge field fluctuations, one can generate statistical anisotropies from the gauge field quantum fluctuations which can be tested observationally.

Observationally,  the imprints of  the gauge fields fluctuations  in primordial curvature perturbation  power spectrum $P_{\calR}$ has the form of  quadrupole anisotropy which is  parametrized  as \cite{Ackerman:2007nb, Pullen:2007tu}
\ba
\label{g*}
\label{g*-def}
P_{\calR}({\bf k}) =   P_{\calR}^{(0)} \left( 1 + g_* (\widehat {\bf n}\cdot  \widehat {\bf k})^2  \right) \, ,
\ea
in which $P_{\calR}^{(0)}$ is the  isotropic power spectrum in the absence of gauge field,  
${\bf k}$ is the mode of interest in Fourier space and $\widehat{\bf n}$ indicates the direction of anisotropy, which in our setup is along the $x$ direction. In this way of parameterization, the parameter $g_{*}$ measures the amplitude of statistical anisotropy. Observational constraints from Planck data implies 
\cite{Kim:2013gka, Planck:2013jfk} $| g_*| \lesssim 10^{-2}$. 

As mentioned above,  it is shown in \cite{Watanabe:2009ct} that
for a broad class of potentials  with the appropriate form of
the coupling $f(\phi)$ the system reaches the attractor regime in which  the electric field energy density reaches a constant and subdominant fraction  of the total energy density. Denoting the fraction of the gauge field  energy density to total energy density by the parameter $R$, the anisotropic correction in primordial curvature perturbation power spectrum is obtained to be 
\ba
\label{g*-eq}
g_* =   -\Big( \frac{48 R}{\epsilon} \Big) N^2 \, ,
\ea
in which  $\epsilon = -\dot H/H^2$ is the  slow-roll parameter and $N$ measures the number of e-folds when the mode of interest  $k$ leaves the horizon towards the end of inflation. Note that  the $N^2$-dependence of the amplitude of  anisotropy is a generic feature  expected from the accumulative IR contributions of the scale-invariant gauge field fluctuations  \cite{Bartolo:2012sd}. Imposing the observational constraints $| g_*| \lesssim 10^{-2}$ one concludes that $R/\epsilon \lesssim 10^{-5}$. The analysis of bispectrum and the trispectrum  were performed in \cite{Abolhasani:2013zya, Shiraishi:2013vja, Shiraishi:2013oqa} in which the amplitude of local-type non-Gaussianity is obtained to be  $f_{NL} \sim g_* N \propto N^3$ with  non-trivial anisotropic shapes.  

\subsection{The EFT action }
\label{EFT action}

Here we present the starting EFT action within the setup of anisotropic inflation.

During the slow-roll inflation with a nearly constant Hubble expansion rate, the quasi de-Sitter background can be parameterized by a single clock i.e. the scalar field $\phi(t)$. Fixing the time slicing by the surface of constant $\phi$, the full four dimensional diffeomorphism invariance is broken to a three dimensional (time dependent) spatial diffeomorphism.  This was employed in \cite{Cheung:2007st} to  write down all the allowed interactions in unitary gauge $\delta \phi=0$ as functions of variables which are 
scalar or  tensor under the remaining three dimensional transformation. Furthermore, the analysis simplifies greatly if one goes to decoupling limit in which the gravitational back-reactions are ignored and the leading interactions are induced from the matter sector. This is particularly useful in non-Gaussianity 
analysis in which it is understood that the gravitational back-reactions does not induce large non-Gaussianities \cite{Maldacena:2002vr}. We also follow this strategy and work in the decoupling limit where all anisotropies are generated from the matter sector. This assumption induces fractional errors at the order of slow-roll  parameters in $g_*$ and $f_{NL}$ which are consistently small.

Our goal is to write down the leading interactions involving the inflaton and the gauge field perturbations.
Going to unitary gauge, one trades $\delta \phi$ fluctuations with quantities such as $\delta g^{00}$ etc.
With the time slicing chosen via $\delta \phi=0$,  the gauge field fluctuations are still left free and we do not have freedom to eliminate them. The situation here is very similar to EFT in multiple fields 
 inflation \cite{Senatore:2010wk} in which the freedom with time diffeomorphism allows one to eliminate only one scalar degrees of freedom. Upon freezing the inflaton fluctuations, the perturbations associated with other field(s) are independent  which can affect the cosmological observables either as an isocurvature field or a heavy field etc. This is the logic we follow in our analysis. 

In an anisotropic inflationary model, the $U(1)$ gauge field breaks the isotropy of the background. However, in the physically relevant limit  of small anisotropy, one expects the background expansion is mainly driven by the isotropic potential as in the single field inflationary models. In this view, the constant time hypersurfaces are determined by the scalar field and the gauge field fluctuations live on these hypersurfaces.  Then, following the logic of EFT and going to decoupling limit, 
the main building blocks in unitary gauge are
\ba
\left\{
  \begin{array}{ll}
    $$\delta g^{00}$$, & \\
    $$X \equiv F_{\mu\nu}F^{\mu\nu}$$, & \\
    $$Y \equiv F_{\mu\nu}\tilde{F}^{\mu\nu}$$, & \\
    $$Z \equiv F^0_{\mu}F^{0\mu}$$ ,  & \\
  \end{array}
\right.
\ea
in which $F_{\mu \nu} = \partial_{\mu } A_{\nu} - \partial_{\nu } A_{\mu} $ is the field strength and
$ \tilde{F}^{\mu\nu} = \epsilon^{\mu \nu \alpha \beta} F_{\alpha \beta} $ is its dual field.

Note that the first term, $\delta g^{00}$, is the usual contribution from the EFT approach while the remaining three operators $X, Y$ and $Z$ represent the contributions of gauge field. The terms 
$X, Y$ and $Z$  are second order in derivatives and are renormalizables while operators constructed from their higher powers are non-renormalizable. Note that the terms $X$ and $Z$ are different in nature. The former is constructed from the contraction of a four tensor while the latter is constructed from the contraction of  a four vector so both are allowed in EFT action.\footnote{We thank S. Mukohyama for pointing out this possibility to us.} $X$ and $Z$ may contribute jointly at the background level, but as we shall see, they source different perturbations. In addition, the appearance of the  term $Y$ is a sign of  parity violation. While we write down the action for general situation including the setup with parity violation, but our main discussions are concerned with scenarios of anisotropic inflation with no parity violation. Finally we comment that in anisotropic inflation models with only $F_{0i}$ being non-zero, we can not have contributions from $F_{0 \mu} \tilde F^{0 \mu}$.

With the above building blocks at hand,  the most general action in unitary gauge $\delta\phi=0$ in the decoupling limit where the gravitational back-reactions are neglected,  is given by 
\ba
{S}
\label{action-UG}
 = \int d^4x\sqrt{-g}&\bigg[& \alpha(t) +B_1(t)\delta g^{00}+\frac{B^2_2(t)}{4}(\delta g^{00})^2-\frac{M_1(t)}{4}\delta X+\frac{M_2(t)}{2}\delta g^{00}\delta X\nonumber \\&&-\frac{M_3(t)}{4}(\delta g^{00})\delta(X)^2
 -\frac{M_4(t)}{4}(\delta g^{00})^2\delta X
 -\frac{N_1(t)}{4}\delta Y +\frac{N_2(t)}{2}\delta g^{00}\delta Y\nonumber\\&&
-\frac{N_3(t)}{4}(\delta g^{00})\delta(Y)^2 -\frac{N_4(t)}{4}(\delta g^{00})^2\delta Y 
-\frac{P_1(t)}{4}\delta Z+\frac{P_2(t)}{2}\delta g^{00}\delta Z\nonumber\\&&-\frac{P_3(t)}{4}(\delta g^{00})\delta(Z)^2
-\frac{P_4(t)}{4}(\delta g^{00})^2\delta Z
\bigg]+... \, .
\ea

The terms $\alpha$ and $B_1$ are determined from the tadpole cancelation at the background level. In particular, $\alpha$ is fixed by the background value of the potential to support inflation while $B \propto \dot H$ in which $H$ is the isotropic Hubble expansion rate.  The couplings $B_2, M_i, N_i$ and $P_i$  are left undetermined from the logic of EFT.  In working with the above general action,  we will keep terms with leading orders of derivatives while terms with higher orders of derivatives are neglected  in low energy. With this logic, the terms containing $M_3, N_3, P_3$ and  higher are higher derivatives compared to $M_1, N_1$ and $P_1$  and are discarded in low energy limit. 

In the small anisotropy limit,  inflation is mainly driven by scalar field which preserves the isotropy of the background. Neglecting the small anisotropies and going to decoupling limit, the background metric is still in the form of  FRW 
\be
ds^2 = -dt^2 + a^2(t) d \vec{x}^2 \, ,
\ee
with the background  Friedmann equations given by
\begin{eqnarray}
\label{tadpoles}
H^2  =  \frac{1}{3 M_{\rm Pl}^2} \big[ B_1(t)+\alpha(t)\big]\, , \quad 
 \dot H + H^2 & =  & -\frac{1}{3 M_{\rm Pl}^2} \big[ 2
B_1(t)-\alpha(t) \big] \;.
\end{eqnarray}
Solving for $\alpha$ and $B_1$, in the small anisotropy limit we have 
\ba
\alpha(t) \simeq  3 M_P^2 H^2,  \quad 
\label{B-back}
B_1(t) \simeq - \epsilon H^2 M_P^2 \, .
\ea

A  key parameter in our discussion is the fraction of the background gauge field energy density to the total energy density, denoted by $R$. In the slow-roll limit when one can neglect the contribution of $B_1$ in total energy density, the gauge field energy density is given by $\frac{1}{2} (M_1 - P_1/2) a^{-2} \dot A_x^2$. Therefore, the fraction of gauge field energy density to total energy density is 
\ba
\label{R-def}
R = \frac{\frac{1}{2}(M_1 - \frac{P_1}{2}) a^{-2} \dot A_x^2}{ 3 M_P^2 H^2} \, .
\ea
As discussed before, we are interested in the attractor limit in which $R$ is small but nearly constant
so the gauge field energy density is a constant fraction of the  total energy density. This criteria determines  the time-dependent forms of $M_1$ and $P_1$.

In addition, from the background Maxwell equation, we have 
\ba
\label{Maxwell-eq}
\partial_t  \left(  (M_1 - \frac{P_1}{2}) a(t) \dot A_x \right) =0 \, .
\ea
Combining this with the above definition of $R$, and assuming that we have reached the attractor limit with $R$ nearly constant, we conclude that $A_x(t) \propto a(t)^{3}$. This is understandable since in order for the gauge field to survive the background expansion, it has to evolve  with appropriate power of the background scale factor. Now with $A_x(t) \propto a(t)^{3}$ a constant value of $R$ can be achieved if one further takes  $M_1 - \frac{P_1}{2} \propto a^{-4}$. Indeed with this choice of 
time dependence for $M_1$ and $P_1$, the gauge field perturbations acquire a scale invariant power spectrum as we shall see in next section.

So far we have kept the terms containing $N_i$ which  induce parity violations in cosmological perturbations.  However, in the analysis below we assume that parity is not violated in primordial universe so we set $N_i=0$ from now on. This is only a matter of simplification. In general, one can extend our analysis to more general situation with parity violating operators turned on.



\subsection{A specific example: Maxwell theory }
\label{Maxwell-sec}

It is instructive to compare our general action (\ref{action-UG})  with the well-studied model of anisotropic inflation  \cite{ Watanabe:2010fh, Bartolo:2012sd} based on Maxwell theory with a $\phi$-dependent gauge kinetic coupling:
\ba
\label{Maxwell}
L_{\mathrm{Maxwell}} = -\frac{f(\phi)^2}{4} F_{\mu \nu} F^{\mu \nu} \, .
\ea
As discussed before, in order for the background electric field energy density to remain a constant sub-dominant fraction of the total energy density we require $f(\phi) \propto a(t)^{-2}$,  yielding 
$ {\dot A_x} = 3 H  \ {A_x}$,  and $ X \propto a(t)^4$  at the background level. Now comparing the Lagrangian  (\ref{Maxwell}) to our general  action (\ref{action-UG}) we have $M_1=f^2\ \propto a^{-4}$  while all other terms $M_2,P_i,...$ are zero.   It is curious  that  the anisotropic inflation based on Maxwell theory is such a simple model compared to
general possibilities encoded in Eq. (\ref{action-UG}). 

In addition, if in the Maxwell  setup we take the potential to be the simple chaotic potential $V(\phi) = \frac{m^2}{2}{\phi^2}$,  then the functional form of $f(\phi)$ yielding the attractor regime is  obtained 
to be \cite {Watanabe:2009ct} 
\ba
\label{f-form0}
f(\phi) = \exp {\left( \frac{c\, \phi^2}{2 M_P^2}  \right)} \, ,
\ea
in which $ c>1$  is a constant. 

Now  the fraction of electric field energy density to the inflaton 
energy, $R$, from Eq. (\ref{R-def}) is obtained to be    $R =I \epsilon/2$ in which the anisotropy parameter $I$ is defined via $I \equiv (c-1)/c$. Calculating the anisotropic power spectrum in this setup \cite{ Watanabe:2010fh,   Emami:2013bk, Abolhasani:2013zya, Abolhasani:2013bpa,  Chen:2014eua, Bartolo:2012sd, Shiraishi:2013vja}
one obtains  $g_* = -24 I N^2 = -48 R N^{2}/\epsilon$  as in Eq. (\ref{g*-eq}). 


\section{The Free Fields and Interactions }
\label{Goldstone sec.}

Having presented the general action in unitary gauge  in Eq. (\ref{action-UG}), we can restore the inflaton fluctuations by performing the transformation 
\ba
\label{pi-mu-transformation}
x^0 \rightarrow {x^0}'=   x^0 + \pi \, ,
\ea
in which the field $\pi$ plays the role of the Goldstone boson. In other words, the Goldstone boson $\pi$ is associated with the breaking of time diffeomorphism which is used to set $\delta \phi=0$ in unitary gauge.
Upon leaving the unitary gauge, we expect to restore the inflaton fluctuations and indeed  the field $\pi$
 encodes the fluctuations of inflaton in an arbitrary coordinate system. 

Upon restoring the Goldstone  bosons $\pi$, the component $g^{00}$ transforms as 
	\ba
	g^{\prime 00}(x^\prime)&=&\frac{\partial x^{\prime 0}}{\partial x^\mu}\frac{\partial x^{\prime 0}}{\partial x^\nu}g^{\mu\nu}(x)\nonumber\\
	&=&\frac{\partial (x^0-\pi(x))}{\partial x^\mu}\frac{\partial (x^0-\pi(x))}{\partial x^\nu}g^{\mu\nu}(x)\nonumber\\
	&=&(\delta_\mu^0-\partial_\mu\pi)(\delta_\nu^0-\partial_\nu\pi)g^{\mu\nu}\nonumber\\
	&=&g^{00}-2\dot\pi g^{00}+\partial_i\pi\partial_j\pi g^{ij}+\dot\pi^2 g^{00} \, ,
	\ea	
and therefore
\be
\delta g^{00}\rightarrow 2\dot\pi+a^{-2}(\pi_{,i})^2-\dot\pi^2.
\ee

As for the gauge field fluctuations we can simply go to Coulomb-radiation gauge  $A_0= \nabla \cdot A=0$ so as usual one ends up with two transverse polarizations of the massless gauge field. One of this perturbation is a scalar while the other one is a vector. More specifically, the scalar part of gauge field perturbations 
has the form $\delta A_{(S)}= (0, \delta A_x,\delta A_y, 0)$. Choosing the wave vector in Fourier space in the form $\bfk = (k_{x}, k_{y}, 0)$, the condition $\nabla \cdot A=0$ implies $k_{x} \delta A_{x} + k_{y} \delta A_{y}=0$. On the other hand, the vector part of gauge field perturbation is given by $\delta A_{(V)}= (0, 0, 0, \delta A_V)$.
In the analysis involving scalar perturbations we neglect the effects of $\delta A_{V}$ but it mixes with the tensor perturbations as we shall see in section \ref{tensor-sec}.

Now expanding $X$ and $Z$ to linear and second order perturbations respectively we obtain 
\ba
\label{X1-eq}
\delta X^{(1)} =- \frac{4}{a^2}\dot{A}_x\delta\dot{A}_x,  \quad   
\delta Z^{(1)}=2a^{-2}\dot{A}_x\delta\dot{A}_x \, ,
\ea
and
\ba
\delta X^{(2)}=\frac{2}{a^2}\Big[-\delta\dot{A}^2_x-\delta\dot{A}^2_y+\frac{1}{a^2}\delta A^2_{x,y}+\frac{1}{a^2}\delta A^2_{y,x} -\frac{2}{a^2}\delta A_{x,y}\delta A_{y,x}\Big] , \quad 
\delta Z^{(2)}=\frac{1}{a^2} [\delta\dot{A}^2_x+\delta\dot{A}^2_y ] .
\ea
Plugging the above values in the action (\ref{action-UG}), the full second order action is obtained to be
\ba
\label{S-total}
{S}=\int d^4x \sqrt{-g} &\Big[& B_1 \left( -\left(\dot{\pi}\right)^2 +a^{-2}\left(\pi_{,i}\right)^2\right) + B_2  \dot{\pi}^2
-\frac{1}{4}M_1\delta X^{(2)}+M_2\dot{\pi}\delta X^{(1)} +\dot{M}_1\pi\delta X^{(1)}
\nonumber \\&&-\frac{1}{4}P_1\delta Z^{(2)}+P_2\dot{\pi}\delta Z^{(1)}+\dot{P}_1\pi\delta Z^{(1)}\Big] \, .
\ea

\subsection{The free fields}

Having obtained the total quadratic action, here we calculate the free actions and the free wave functions for  $\pi$ and 
$\delta A_i$ fluctuations. 

The free action of $\pi$ from Eq. (\ref{S-total}) is given by 
\begin{align}
\label{pi-action}
{S}^{(\pi)}_2=\int d^4x\sqrt{-g}\, & (-B_1) \left[\left(1-\frac{B_2}{B_1}\right) \dot{\pi}^2 - a^{-2}\left(\pi_{,i}\right)^2\right] \, .
\end{align}
Note that $B_1 \propto \dot H  <0$ so the kinetic energy has the correct sign.

The free wave function of $\pi$ with the Minkowski initial conditions deep inside the horizon  is
\begin{equation}
\label{pi-wave}
\pi(k)=\frac{H}{2k^{3/2}\sqrt{c_s|B_1|}}(1+ikc_s\tau)e^{-ikc_s\tau},
\end{equation}
where we have defined the sound speed of $\pi$ fluctuations, $c_s$, as
\begin{equation}
\label{cs-eq}
{c_s^{-2}=1-\frac{B_2}{B_1}}.
\end{equation}
Note that, the wave function Eq. (\ref{pi-wave}) differs from the canonically normalized wave function by a factor of $1/\sqrt 2$.

Within the approach of EFT the coefficient $B_{2}$, containing the operator 
$(\delta {g^{00}})^{2}$, controls the sound speed of $\pi$ fluctuations  \cite{Cheung:2007st}. This operator can arise for example in the models 
with non-trivial kinetic energy such as in k-inflation \cite{ArmendarizPicon:1999rj, Garriga:1999vw}
or DBI inflation  \cite{Alishahiha:2004eh}.  Interestingly, within the EFT approach, we can extend the DBI-type model to anisotropic inflation with gauge fields. As a motivation, this scenario may arise within the setup of  string theory in which a D3 brane containing $U(1)$ gauge fields moves ultra relativistically inside an AdS throat \cite{Dimopoulos:2011pe}.

The free wave function of $\delta A_i$ comes from the following contributions 
\ba
{S}^{(\delta A)}_2=\int d^4x\sqrt{-g} \left(-\frac{1}{4}M_1\delta X^{(2)}-\frac{1}{4}P_1\delta Z^{(2)}\right) \, .
\ea
Using the relation
\begin{equation}
\epsilon ^{ijk}\delta A_{j,k}\delta \dot{A}_i=\frac{1}{2}\left[ \partial _0 \left(\epsilon ^{ijk} \delta A_i \delta A_{j,k}\right) - \partial _k \left( \epsilon ^{ijk} \delta A_i \delta \dot{A}_j\right)\right],
\end{equation}
the  free field action for $\delta A_i$ fluctuations  is given by 
\begin{align}
{S}_2^{(\delta A)}=&\int d^4x \sqrt{-g} \frac{1}{2a^2} \left[(M_1-\frac{P_1}{2})\left(\delta \dot{A}_{i}\right)^2 -M_1 a^{-2} \left( \epsilon_{ijk} \delta A^{i, j}\right)^2\right]  \, .
\label{action-A}
\end{align}

So far we have not specified the form of the time dependence of parameters $M_1$ and $P_1$.
However, as we discussed at the end of section \ref{EFT action}, in order to reach the attractor regime 
in which the gauge field energy density becomes a constant fraction of the total energy density, we require $A_x \propto a(t)^3$ and $M_1 - \frac{P_1}{2} \propto a(t)^{-4}$. With these scalings, the parameter $R$, measuring the fraction of the gauge field energy density to total energy density, reaches 
a constant value. We also show below that with this choice of scaling for $M_1$ and $P_1$, the gauge field fluctuations acquire a scale invariant power spectrum. 

To simplify the notation, we absorb the time scaling of $M_1 - \frac{P_1}{2}$ and $A_x(t)$ via
\ba
\label{bar-values}
\tilde{M}_1 \equiv M_1 - \frac{P_1}{2} \equiv \overline M_1 a^{-4} , \quad \quad  \dot A_x(t)  \equiv \overline A a^{3} \, ,
\ea
in which $\overline M_1$ and $\overline A$ are constants.  With these definitions, the parameter $R$ from  Eq. (\ref{R-def}) simplifies to
\ba
\label{R-value}
R = \frac{ \overline M_1 \overline A^2}{6 M_P^2 H^{2}} \, .
\ea
 We will use this relation to eliminate the combination $\overline M_1 \overline A^2$ in favor of the physical parameter $R$.

 Now we go back to the analysis of gauge field fluctuations  in action (\ref{action-A}). First, note that  it will be very convenient to decompose the gauge field fluctuations  in terms of its polarization base $\epsilon ^s_i(k)$ in Fourier space
\begin{equation}
\delta A_{i}=\sum _s \delta A^{\,  (s)}(k,t)\epsilon ^s_i(k) \, .
\end{equation}
The polarization vector can have either the   linear polarization form with $s=1,2$ or the circular (helicity) polarization form  with $s =\pm$. 

Now imposing the Minkowski initial condition for the gauge field fluctuations  deep inside the horizon we obtain
\begin{equation}
\label{A-wave}
\delta A^{(s)}_i =\frac{1}{{k^{3/2}\sqrt{2c_v\overline{M}_1}}H\tau^3}(1+ikc_v\tau )e^{-ikc_v\tau} \, ,
\end{equation}
in which $c_v$ represents the speed of gauge field fluctuations 
\begin{equation}
\label{cv-eq}
c^2_v=\frac{M_1}{\tilde M_1} \, , 
\end{equation}
with  $\tilde M_1$ defined in Eq. (\ref{bar-values}).  Note that in simple Maxwell theory $c_v=1$ while in our general setup it can be different than unity. 

So far we have only been able to determine the time dependence  of the combination $\tilde M_1= M_1 - \frac{P_1}{2} $. Now in the expression for $c_v$ we see that the degeneracy with $M_1$ and $P_1$ is broken and they appear differently 
in $c_v$. For example,  in simple Maxwell theory we have $c_{v}=1$ so from this requirement we conclude that $P_{1}=0$ and only $M_{1}$ is non-zero in Maxwell theory.

It is reasonable to assume that $c_v$ is constant. This also fixes the scaling of $M_1$ separately 
to be $M_1 \propto a^{-4}$ so at the end we conclude that $P_1 \propto M_1 \propto a^{-4}$.


\section{Anisotropic Power Spectrum}
\label{power}
Having obtained the free wave functions, we can proceed to calculate the leading interactions between 
$\pi$ and $\delta A_i$ fluctuations which source the anisotropies. 

The leading interactions  involving $\pi$ and $\delta A_i$ fluctuations  from Eq. (\ref{S-total}) 
are given by
\begin{align}
{S^{(\pi \delta A)}} = \int d^4x\sqrt{-g}\left[\dot{\tilde{M}}_1a^{-2}\dot{A}_x\pi\delta\dot{A}_x-4\tilde{M}_2(t)a^{-2}\dot{A}_x\dot{\pi}\delta\dot{A}_x \right] \, ,
\end{align}
in which, similar to $\tilde M_1$, we have defined  $\tilde M_2 \equiv M_2-\frac{P_2}{2} $. 

To go further, we have to determine the scaling of $\tilde M_2(t)$ with time.  Following the same logic yielding the scaling of $\tilde M_1$  and noting that $M_{2}$ and $P_{2}$  are generated 
respectively from the higher order interactions involving $M_{1}$ and $P_{1}$, 
we also conclude that $\tilde M_2\propto  \tilde M_1\propto a^{-4}$. 
With these  scalings of $\tilde M_i$,  the second order interaction Lagrangian written in the conformal time $d \tau = dt/a(t)$ is
\ba
{S}^{(\pi \delta A)}=\int d\tau d^3 x \left(L_1+L_2\right),
\ea
with
\begin{align}\label{2-interaction}
L_1=4H \overline{A}\,  \overline{ M}_1\pi \delta{A_{x}'} \, , \quad \quad 
L_2=4 \overline{A} a^{-1}\overline{ M}_2\pi^{\prime}\delta{A_{x}'} \, ,
\end{align}
in which a prime indicates the derivative with respect to conformal time. Also note that similar to $\tilde M_1$, we have absorbed the scaling of $\tilde M_2$ such that $\tilde M_2 \equiv \overline M_2 a^{-4}$ with $\overline M_2$ being a constant. 

\begin{figure}[t!]
  \hspace{1cm}
  \includegraphics[width=0.8\linewidth]{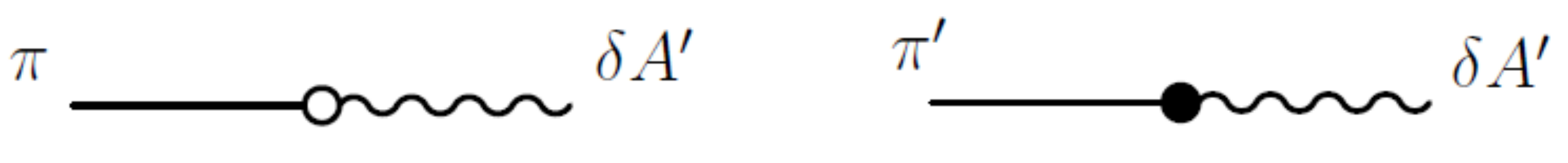}
  \caption{ The exchange vertices from two Hamiltonians in Eq. (\ref{2-interaction}). The solid (wavy) line represents the inflaton (gauge field) fluctuations while the empty and the filled circles indicate the two different interactions in Eq. (\ref{2-interaction}). 
  }
  \label{scalar-vertex}
\end{figure}

To use the perturbative in-in formalism, we need to obtain the interaction Hamiltonians $H_{1}$ and $H_{2}$
constructed respectively from $L_{1}$ and $L_{2}$. For $L_{1}$ we simply have $H_{1} =- L_{1}$. However, for $H_{2}$ we can not simply set $H_{2} =-L_{2}$. This is because there is derivative coupling involving $\pi'$ so we have to construct $H_{2}$ from its usual definition involving the conjugate momentum. We obtain 
\ba
\label{2-interaction}
H_1=-4H \overline{A}\,  \overline{M}_1\pi\delta A_{x}' \, , \quad  \quad 
H_2=-4 \overline{A}\, \overline{M}_2\Big(1+\frac{8\overline{A}^2\overline{ M}_2^2}{\overline{ M}_1B_1c_s^{-2}}\Big) a^{-1}\pi^{\prime}\delta{A_{x}'} \, .
\ea
From the form of $H_{2}$ we see that the correction in $H_{2}$ ( in not using  $H_{2}= -L_{2}$) is 
at the order of $\overline A^{2} \propto R $. Therefore,  in the limit of small anisotropy $R \ll 1$,  one can safely neglect these corrections. This is equivalent to simply setting $H_{2}= -L_{2}$ to leading order in anisotropy.   In Fig. \ref{scalar-vertex}  a schematic view of the two mixing interactions in Eq. (\ref{2-interaction}) is presented.

Using the standard in-in formalism,  the anisotropic corrections to $\langle \pi \pi \rangle $ power spectrum is calculated via \cite{Emami:2013bk, Chen:2014eua}
\ba\label{Corr}
\delta P_{ij}=- \int_{-\infty}^{\tau_e}d\tau_1\int_{-\infty}^{\tau_1}d\tau_2\Big\langle\Big[H_i(\tau_2), \, \big[H_j(\tau_1),\pi^{\ast}(\tau_e)\pi^{\ast}(\tau_e)\big]\Big]\Big\rangle,
\ea
where $\tau_e$ represents the time of end of inflation and $H_i$ and $H_j$ are either $H_1$ or $H_2$ given in Eq. (\ref{2-interaction}). However, note that the observable comoving curvature perturbations $\mathcal{R}$ is related to $\pi$ via
\ba\label{Curv}
\mathcal{R}=-H\pi+O(\pi^2),
\ea
so when calculating the curvature perturbation power spectrum $P_{\calR}$, we simply multiply $\delta P_{ij}$ by a  factor of $H^{2}$.

\begin{figure}[t!]
  \hspace{0cm}
  \includegraphics[width=1\linewidth]{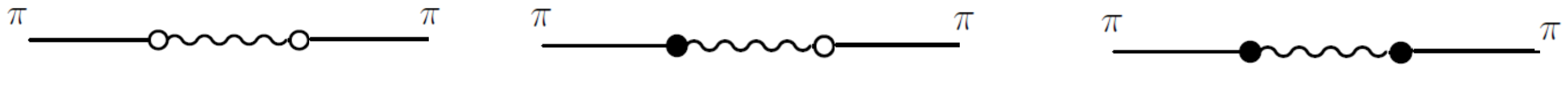}
  \caption{ The three different anisotropic corrections to scalar two point functions from exchanges vertices of  Hamiltonians (\ref{2-interaction}). }
  \label{scalar-power-mix}
\end{figure}

As seen from Fig. \ref{scalar-power-mix} there are three different contributions in $\delta P_{ij}$ depending on how one uses the two exchange vertices from the Hamiltonians  (\ref{2-interaction}) inside the nested integrals.  Using the wave functions for $\pi$ and $\delta A_i$ fluctuations 
respectively given in (\ref{pi-wave}) and (\ref{A-wave}), we obtain
\ba\label{power-aniso}
\delta P_{11}&=&-\frac{H^2 c^4_s N^2}{16B_1^2 k^3c_v\overline{ M}_1}\left(-4H \overline{A}\,  \overline{ M}_1\right)^2 \sum _s |\epsilon ^s_1(k)|^2,\\
\delta P_{12}= \delta P_{21}
&=&\frac{-3H^3 c^4_s N^2}{16B_1^2 k^3c_v\overline{ M}_1}\left(-4H \overline{A}\,  \overline{ M}_1\right)
\Big(-4 \overline{A}\, \overline{ M}_2\big[1+\frac{8\overline{A}^2\overline{ M}_2^2}{\overline{ M}_1B_1c_s^{-2}}\big]\Big) \sum _s |\epsilon ^s_1(k)|^2,\\
\delta P_{22}&=&-\frac{9H^4 c_s^4 N^2}{16B_1^2 k^3\overline{M}_1c_v}\Big(-4 \overline{A}\, \overline{ M}_2\big[1+\frac{8\overline{A}^2\overline{ M}_2^2}{\overline{ M}_1B_1c_s^{-2}}\big]\Big)^2 \sum _s |\epsilon ^s_1(k)|^2,
\ea
in which $N$ represents the total number of e-folds when the mode of interest leaves the horizon till the end of inflation.  Note that  for illustration  we have kept the sub-leading terms in $H_{2}$ containing higher powers of $\overline A^{2} $ which yields the higher corrections in $\delta P_{2 i}$ as given above.  
However in the following analysis, we safely ignore these higher corrections  in anisotropy power spectrum.  Finally, the total anisotropic power spectrum $\delta P_{\mathrm{total}}$ is obtained from the sum of the above four contributions.

So far we have not specified the polarization vectors. To simplify the result we choose the wave number such that it has the symmetry in the $yz$ plane  and 
\begin{equation}
\mathbf{k}=k\left(\cos \theta ,\sin \theta ,0\right) \, ,
\end{equation} 
where $\theta $ is the angle between the wave number and the preferred direction $\hat {\bf n}$,  i.e. $\cos \theta = \widehat {\bf  k} \cdot \widehat {\bf n}$ which in  our case  $\widehat {\bf n}$ is along the $x$ direction.  The polarization base can be  either the linear base or the helicity base.  Starting with  the linear base
\ba
\label{linear-base}
\epsilon^{(1)} = (-\sin \theta, \cos \theta, 0) \quad ,\quad \epsilon^{(2)} = (0, 0, 1)  \, ,
\ea
 the helicity base can be written in terms of the linear base as follows 
\ba
\label{circular-base}
\epsilon^{(+)} = \frac{i}{\sqrt2} ( \epsilon^{(1)} + i \epsilon^{(2)}) \quad , \quad
\epsilon^{(-)} = \frac{-i}{\sqrt2} ( \epsilon^{(1)} - i \epsilon^{(2)}) \, .
\ea
Using either base we obtain $\sum _s |\epsilon ^s_1(k)|^2 = \sin^2 \theta$. 

We are interested in  fractional change in power spectrum, $\delta P_{\mathrm{total} }/ P_{\pi}^{(0)}$, in which  $P_{\pi}^{(0)}$ is the isotropic power spectrum from $\pi$ field, 
\begin{equation}
P_{\pi}^{(0)}=\frac{H^2}{8 |B _1| k^3c_s},
\end{equation}
with  $B _1$ obtained  from the tadpoles cancellation in Eq. (\ref{B-back}).  Now adding the four contributions of $\delta P_{ij}$ obtained above, and discarding the sub-leading powers of $\overline A$ in $\delta P_{2 i}$ as discussed before,  the total fractional change in power spectrum is obtained to be 
 \ba
 \label{deltaP-ratio1}
\frac{\delta P_{\mathrm{total}}}{P_{\pi}^{(0)}} = \frac{8H^2c^5_s  \overline M_1 \, \overline{A}^2}{B_1c_v }  \left( 1+ \frac{3 \overline M_{2}}{\overline M_{1}} \right)^{2}    N^2 \sin^2 \theta  \, .
\ea 

Comparing the above expression with the amplitude of quadrupole anisotropy $g_*$ defined in Eq. (\ref{g*-def}) yields  
\ba
\label{g*-total}
g_* = -\frac{48 R c^5_s N^{2}}{\epsilon \, c_v}  \left( 1+ \frac{3 \overline M_{2}}{\overline M_{1}} \right)^{2} \, ,
\ea
in which we have used Eq. (\ref{R-value}) to eliminate $ \overline M_1 \, \overline A^2$ in favor of $R$. 

We note the curious conclusion that $g_{*}$ is always negative. 
Also note  the overall $N^2$ dependence which is common in models of anisotropic inflation from the accumulations of  IR gauge field fluctuations \cite{Bartolo:2012sd}. In addition, there is the  factor  $c_s^5/c_v$ coming from the non-trivial speeds of scalar field  and gauge field fluctuations. The above result also agrees with that of \cite{Abolhasani:2015cve}. However, the factor $c_{v}$ in the denominator above is missing in the analysis of \cite{Abolhasani:2015cve}. This is because  the effects of the terms containing $P_{i}$ are missing in the analysis of \cite{Abolhasani:2015cve}.  As we have seen, the only effect of $P_{1}$ is to break the degeneracies between $M_{1} $ and $P_{1}$ yielding a non-trivial value for $c_v$ 
as given in Eq. (\ref{cv-eq}).

As an example, we compare the above result with the special case of anisotropic inflation in Maxwell theory as reviewed in section \ref{Maxwell-sec} with only $M_1$ being non-zero while all other operators are zero.  In this case 
$g_{\ast}$ is obtained to be 
\be
g_* =-\frac{48 R }{\epsilon} c_{s}^{5} N^{2} \, .
\ee
In addition, if we assume the scalar field has the conventional kinetic energy with $c_{s}=1$, the above 
value of $g_{*}$  agrees  with Eq. (\ref{g*-eq}) obtained in previous works of anisotropic inflation 
\cite{ Watanabe:2010fh,  Emami:2013bk, Abolhasani:2013zya,  Bartolo:2012sd}.

\section{Bispectrum}
\label{bispectrum}

Now we look at the bispectrum analysis.  It is in the bispectrum analysis where the EFT approach 
shows its strength. In scenarios of inflation with non-trivial matter sector, such as in  models of  anisotropic inflation  or in models with non-trivial kinetic energy  \cite{ArmendarizPicon:1999rj, Garriga:1999vw, Alishahiha:2004eh}, the leading non-Gaussianities are sourced from the interactions  in the matter sector. Therefore, one can safely go to decoupling limit, i.e. 
neglect the gravitational back-reactions, and obtain the leading non-Gaussianities from the matter sector. This is exactly the logic we have followed  in writing  our starting action Eq. (\ref{action-UG}). 

The shapes of non-Gaussianities of course depends on the form of interactions. 
In simple Maxwell theory with $\phi$-dependent gauge kinetic coupling, the leading three point interaction has the form $f(\phi) f(\phi)_{,\phi} \delta \phi \delta \dot A_{i}^{2}$, i.e. one $\pi$ couples to two gauge field fluctuations. The interactions containing  $\pi^{2}$ with one $\delta A_{i}$ fluctuation are suppressed. This is because these interaction are accompanied with an  additional background factor $\dot A_{x} \propto \sqrt{R} $ which is very small in the limit of small anisotropies.  

In our EFT approach the term $ M_{1}$ is equivalent to the gauge kinetic coupling $f(\phi)^{2}$ in Maxwell theory. However, we have the new interactions $\delta g^{00} \delta X$  and $\delta g^{00} \delta Z$ which are  beyond the Maxwell theory. As discussed before, these interactions  may arise for example from the extension of DBI model for a mobile brane with a $U(1)$ gauge field on its world volume. It is interesting that our EFT approach naturally incorporate these scenarios \cite{Dimopoulos:2011pe}.  We expect to obtain new non-Gaussianities from these operators.  

Note that in our analysis below we calculate  the anisotropic contribution  to $\langle \pi^{3} \rangle$. On top of this, we have the usual isotropic bispectrum originated from the gravitational back-reactions of $\pi$ fluctuations  which induce non-Gaussianity at the order $f_{NL}^{\mathrm{iso}} \sim O(\epsilon)$ \cite{Maldacena:2002vr}. However,  we are interested in large non-Gaussianity obtained from the interactions involving $\pi$ and $\delta A_{i}$ which is anisotropic in nature.    

With these discussions in mind, the cubic interactions from our action (\ref{action-UG}) is obtained to be   
\begin{eqnarray}\label{Third-ac}
S^{(3)}_{int}=\int d^4 x\sqrt{-g} &\Bigg[ &-\frac{\dot{M}_1}{4}\pi\delta X^{(2)}-\frac{\ddot{M}_1}{8}\pi^2\delta X^{(1)}+M_2\dot{\pi}\delta X^{(2)} +\left(M_2+M_4\right)\pi\dot{\pi}\delta X^{(1)}
\nonumber \\
&&+\left(\pi_{,\mu}\pi^{,\mu}\frac{M_2}{2}-6\frac{M_4}{4}\dot{\pi}^2\right)X^{(1)}
-\frac{\dot{P}_1}{4}\pi\delta Z^{(2)}-\frac{\ddot{P}_1}{8}\pi^2\delta Z^{(1)}+P_2\dot{\pi}\delta Z^{(2)}
\nonumber \\
&&+\left(P_2+P_4\right)\pi\dot{\pi}\delta Z^{(1)}+\left(\pi_{,\mu}\pi^{,\mu}\frac{P_2}{2}-6\frac{P_4}{4}\dot{\pi}^2\right) \delta Z^{(1)}\Bigg ] \, .
\end{eqnarray}
Many of the terms above are sub-leading as follows. As can be seen from Eq. (\ref{X1-eq}), the terms involving $ \delta X^{(1)}$ and $\delta Z^{(1)}$
contain the background gauge field $\dot A_{x} \propto \sqrt R $ 
which is small in the limit of small anisotropy. Therefore, the leading interactions come from the second order  gauge field fluctuations $\delta X^{(2)}$ and $ \delta Z^{(2)} $ and the first order of  inflaton field perturbation.

\begin{figure}[t!]
  \hspace{1.5cm}
  \includegraphics[width=0.8\linewidth]{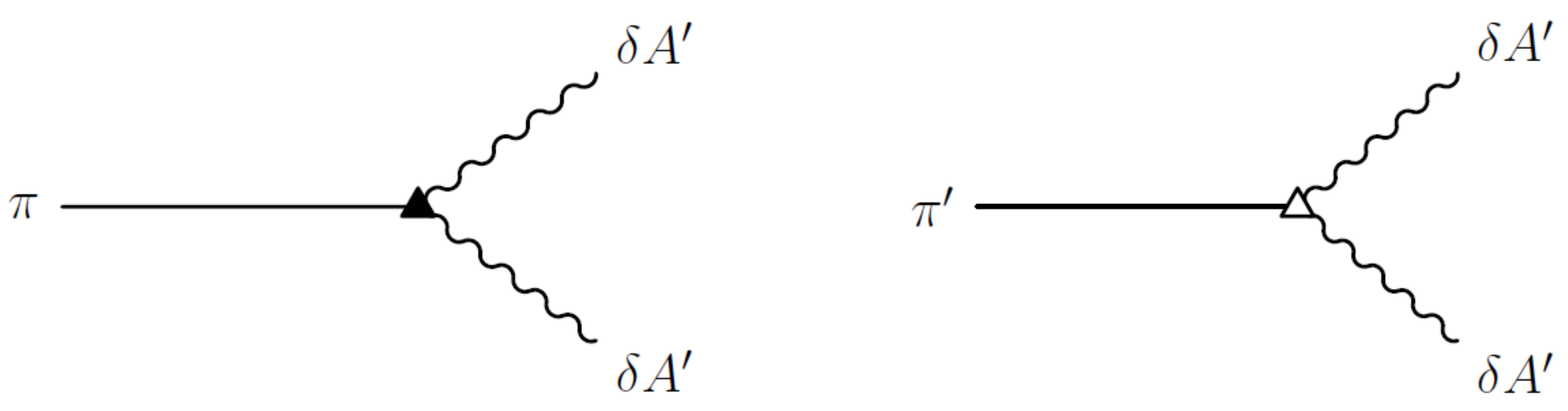}
  \caption{ The three point interactions from  Eq. (\ref{3-interaction}) with the filled and empty triangles representing the two different types of  exchange vertices. }
  \label{scalar-cubic}
\end{figure}

Correspondingly, the leading  cubic interactions Lagrangians in conformal time are 
\be
\label{3-interaction}
L_1^{(3)}=2a^{-4}H\overline{M}_1\pi\delta A^{\prime 2}, \quad \quad 
L_2^{(3)}= 2a^{-5}\overline{M}_2\pi^{\prime}\delta A^{\prime 2} \, .
\ee
From the above Lagrangians one can construct the cubic interaction Hamiltonians. As discussed in the case of power spectrum, to leading order in $R$, we can safely take the cubic action to be $H^{(3)} = -L^{(3)}$ in which $L^{(3)}= L_1^{(3)} + L_2^{(3)}$ as given above. The above  cubic interactions are schematically presented in Fig. \ref{scalar-cubic}.

Neglecting the isotropic contribution in bispectrum which is small in the slow-roll limit,  the leading anisotropic contributions to bispectrum is given by the following nested integrals  \cite{Weinberg:2005vy, Abolhasani:2013zya, Shiraishi:2013vja, Shiraishi:2013oqa}
\be\label{Bispec}
\Big \langle  \pi (\bfk_1) \pi(\bfk_2) \pi (\bfk_3) \Big \rangle_{ijk} &=& 
i\int_{-\infty}^{\tau_e}d\tau_1\int_{-\infty}^{\tau_1}d\tau_2\int_{-\infty}^{\tau_2}d\tau_3\Big\langle  \Big[ H_i({\tau _3}), \Big[ H_j({\tau _2}), \Big[H_k({\tau _1}) ,{\pi^3(\tau _e)}\Big ]  \Big] \Big]  \Big\rangle \nonumber\\
 & \equiv &(2\pi)^3\delta^3(\mathbf{k}_1+\mathbf{k}_2+\mathbf{k}_3) B_{ijk} (\bfk_1, \bfk_2, \bfk_3) \, .
\ee 
In order to calculate the anisotropic bispectrum from the above integrals, one of $H_i({\tau _j})$ should be  the cubic Hamiltonian $H_i^{(3)}$ while the other two  $H_j({\tau _k})$ should be  the quadratic interactions $H_i^{(2)}$ given in Eq. (\ref{2-interaction}). There are three possibilities to put $H_i^{(3)}$. However, these different possibilities are equivalent so it is enough to calculate one of the three terms and multiply the result by a factor of three \cite{Emami:2014tpa}.
In what follows we denote the contributions to the bispectrum from different terms by $B_{ij(k)}$ in which the place of cubic Hamiltonian $H_i^{(3)}$ is shown by $(k)$ while $i j$ represents the location of 
quadratic Hamiltonian from Eq. (\ref{2-interaction}). With this notation, we have 
\ba
\label{Bi}
&B_{11(1)}&= \frac{-6 c_s^6  H^9 \,\overline{M}_1 \overline{A}^2 }{c_v^2 B_1^3 } N_{k_1}  N_{k_2}N_{k_3} 
\big( \frac{C(\mathbf{k}_2,\mathbf{k}_3)}{k_2^3k_3^3}+2  \mathrm{c.p.} \big)   \\
&B_{22(1)}&= \frac{-54 c_s^6  H^9 \,\overline{M}_2^2  \overline A^2}{c_v^2 B_1^3 \overline M_1} N_{k_1}  N_{k_2}N_{k_3}  
\big( \frac{C(\mathbf{k}_2,\mathbf{k}_3)}{k_2^3k_3^3}+2  \mathrm{c.p.}\big)  
\\
&B_{12(1)}&=B_{21(1)}= \frac{-36 c_s^6  H^9 \,\overline{M}_2  \overline A}{c_v^2 B_1^3} N_{k_1}  N_{k_2}N_{k_3}
\big( \frac{C(\mathbf{k}_2,\mathbf{k}_3)}{k_2^3k_3^3}+2  \mathrm{c.p.}\big) \, ,
\ea
and  
\ba
&B_{11(2)}&=  \frac{-18 c_s^6  H^9 \,\overline{M}_2 \overline{A}^2 }{c_v^2 B_1^3 } N_{k_1}  N_{k_2}N_{k_3} 
\Big( \frac{C(\mathbf{k}_2,\mathbf{k}_3)}{k_2^3k_3^3}+2 \mathrm{c.p.}\Big)   \\
&B_{22(2)}&= \frac{-162 c_s^6  H^9 \,\overline{M}_2^3  \overline A^2}{c_v^2 B_1^3 \overline M_1^2} N_{k_1}  N_{k_2}N_{k_3}  
\Big( \frac{C(\mathbf{k}_2,\mathbf{k}_3)}{k_2^3k_3^3}+2  \mathrm{c.p.}\Big)    \\
&B_{12(2)}&= B_{21(2)}= \frac{-108 c_s^6  H^9 \,\overline{M}_2^2  \overline A}{c_v^2 B_1^3 \overline{M}_1} N_{k_1}  N_{k_2}N_{k_3}
\Big( \frac{C(\mathbf{k}_2,\mathbf{k}_3)}{k_2^3k_3^3}+2  \mathrm{c.p.}\Big)  \, ,
\ea
in which $\mathrm{c.p.}$ represents the cyclic permutations. 
Note that  in obtaining the above results, we have neglected the sub-leading terms containing
$\overline A^4  \sim  {\cal O}( R^2)$. These sub-leading terms arise from the higher order corrections in 
quadratic interaction Hamiltonian $H_2$ in Eq. (\ref{2-interaction}). 

In the above expressions, $N_{k_i}$ represents the number of e-fold when the mode $\bfk_i$ leaves the horizon. In practice we take $N_{k_i} \sim N\sim 60$.   In addition,  in the above expressions, we have defined the anisotropic shape function $C(\mathbf{k}_2,\mathbf{k}_3)$ which results in from various contractions of gauge field fluctuations given by \cite{ Bartolo:2012sd, Abolhasani:2013zya, Shiraishi:2013vja, Shiraishi:2013oqa}
\ba
C(\mathbf{k}_2,\mathbf{k}_3) \equiv 1- (\widehat {\mathbf{n}}. \widehat{\mathbf{k}}_2)^2 -  (\widehat{\mathbf{n}}. \widehat{\mathbf{k}}_3)^2 + 
(\widehat{\mathbf{n}}. \widehat{\mathbf{k}}_2) \,  (\widehat{\mathbf{n}}. \widehat{\mathbf{k}}_3)  \, (\widehat{\mathbf{k}}_2 . \widehat{\mathbf{k}}_3) \, , 
\ea
with similar  definition for $C(\mathbf{k}_1,\mathbf{k}_3)$ and $C(\mathbf{k}_1,\mathbf{k}_2)$. 
We remind that $\widehat {\mathbf{n}}$ represents the orientation of anisotropy, which in our example
is the $x$-direction. 

Combining all contributions from $B_{ij (k)}$, eliminating the factor $\overline M_1 \overline A^2$ in favor of the anisotropy parameter $R$ from Eq. (\ref{R-value}),  and using Eq. (\ref{B-back}) to eliminate $B_1$   from the tadpole cancellation,  the total anisotropic  bispectrum is given by
\ba
\label{B-total}
B_{\text{tot}}(\bfk_1, \bfk_2, \bfk_3)=36\frac{c_s^6 H^5 R }{\epsilon^3 c^2_v M_P^4} N_{k_1}N_{k_2}N_{k_3}
\bigg( 1  + 27\frac{\overline M_2^3}{\overline M_1^3}+45\frac{\overline M_2^2}{\overline M_1^2}+15 \frac{\overline M_2}{{\overline M_1}}\bigg)
 \Big( \frac{C(\mathbf{k}_2,\mathbf{k}_3)}{k_2^3k_3^3}+2  \mathrm{c.p.}\Big) .
\ea
It is interesting that non-trivial combinations of $c_v, c_s$ and the fraction  $\overline M_2/\overline M_1$
appear in the anisotropic bispectrum. In principle, in conjugation with the anisotropic power spectrum Eq. (\ref{g*-total}), one can use the shape and the amplitude of the bispectrum to put constraints on the parameters of the  EFT of anisotropic inflation.

For the special   case of Maxwell theory with only $M_1$ being non-zero and taking $c_s=1$, the above expression simplifies to
\ba
B_{\text{tot}}(\bfk_1, \bfk_2, \bfk_3)= \frac{576 R}{\epsilon}  N_{k_1}N_{k_2}N_{k_3} \Big(4 P_\pi^{(0)} (\bfk_1)  P_\pi^{(0)} (\bfk_2)  C(\mathbf{k}_1,\mathbf{k}_2)+2 \mathrm{c.p.} \Big)  \, ,
\ea
which agrees with the result of \cite{ Bartolo:2012sd, Abolhasani:2013zya, Shiraishi:2013vja, Shiraishi:2013oqa}. Note that the additional factor $4$ comes from the fact that our field $\pi$ is off by a factor $1/\sqrt 2$ from the canonically normalized field. 

If one extends the notion of non-Gaussianity parameter $f_{NL}$ to our anisotropic setup, from Eq. (\ref{B-total}) we conclude that $f_{NL} \sim 10 |g_*|  N$. Taking the observational bound $|g_*| \lesssim 10^{-2}$
 and assuming $N= 60$ we obtain  $f_{NL} \lesssim 6$. This value of $f_{NL}$ seems large enough to be detected by current or upcoming observations. Note that the  bispectrum in Eq. (\ref{B-total})  has a specific anisotropic shape controlled by the shape function $C(\mathbf{k}_i,\mathbf{k}_j) $ so the imprints of this shape on CMB maps will be somewhat different than the standard local-type or equilateral-type 
  non-Gaussianities.

\section{Gravitational Waves}
\label{tensor-sec}

In this section we study the tensor perturbations in our EFT approach. One interesting aspect of studying tensor perturbations within the setup of anisotropic inflation is that there will be cross correlation between the scalar and tensor sectors. This is because the background is anisotropic so the usual decoupling of  scalar, vector and tensor perturbations in two point correlations does not hold. Our aim here is to study the scalar-tensor cross correlation and also the anisotropies in tensor power spectrum from the EFT approach. 

Before proceeding, there is one important point to be clarified. As emphasized in writing our starting action in unitary gauge Eq. (\ref{action-UG}), we work in the decoupling limit so the gravitational back-reactions are neglected. This amounts to neglecting the slow-roll corrections in power spectrum and specially in non-Gaussianity analysis. It may sound confusing how our treatment of tensor perturbations and effects such as scalar-tensor cross correlation is consistent with the assumption of decoupling limit. The point is that we are looking for interactions in matter sector which directly mix the tensor perturbations with the gauge field fluctuations and scalar perturbations. These types of interactions come from operators like $M_1$ and $P_1$ etc. We do not need to take into account the sub-leading slow-roll corrections, say from the potential, to obtain the scalar-tensor cross correlation. In this view, even within the assumption of decoupling limit, there still are dominant direct interactions which play non-trivial effects for mixing tensor-scalar and tensor-gauge field perturbations.

In the limit of small anisotropy where one can neglect the differences in background scale factors in three different directions,  the tensor perturbations in metric are \cite{Chen:2014eua}
\ba
\label{dynamical metric}
ds^2 = a(\tau)^2 \left(-d\tau^2 + \left(\delta_{ij}+ h_{ij} \right) dx^i dx^j \right ) \, .
\ea
The perturbations $h_{ij} $ are subject to the transverse and traceless conditions  $\partial_i h_{ij} =h_{ii}= 0$  in which the repeated indices are summed over.  

Now decomposing $h_{ij}$ into its polarization base $e_{ij}^{(s)}(\bfk)$ in Fourier space, from  the traceless  and transverse conditions we obtain 
\ba
e_{i i}^{(s)}(\bfk) = 0 \quad , \quad  k_j e_{i j}^{(s)}(\bfk) = 0  \, ,
\ea
in which $s= \times, +$ represent the two independent polarization of tensor perturbations. We choose the following normalization
\ba
e^{(s)}_{ij}(\mathbf{k}) e^{*(s')}_{ij}(\mathbf{k}) = \delta_{ss'} \, ,
\ea
where $*$ represents the complex conjugation. In addition the relation $e^{(s)}_{ij}(\mathbf{k}) = e^{*(s)}_{ij}(\mathbf{-k})$ holds. Note that  in this section,   the polarization index $s$ is only for tensor perturbations and does not apply to gauge field perturbations. 

The  quantum operators $\widehat{h}_{ij}(\mathbf{k},\tau) $ 
are decomposed in terms  of the annihilation and creation operators as usual via
\ba
\label{tensor}
\widehat{h}_{ij}(\mathbf{k},\tau) = \sum _{s=+,\times}  \widehat{h}_{s}(\mathbf{k},\tau)
e_{ij}^{(s)}(\bfk)
\quad , \quad \widehat{h}_{s}(\mathbf{k},\tau)=
h_{s}(k, \tau)a_{s}(\mathbf{k})+ h^{*}_{s}(k, \tau)a^{\dag}_{s}(-\mathbf{k}) \, ,
\ea
with  the usual commutation relations  $ [ a_{s}(\bfk),    a_{s}^\dagger  (\bfk')] = \delta_{s s'} \delta^{(3)} (\bfk-\bfk' )$.

With our convention of the  the wave vector  $\mathbf{k} = k(\cos{\theta}, \sin{\theta}, 0)$,  the polarizations
$e^{+}_{ij}(\mathbf{k})$ and $e^{\times}_{ij}(\mathbf{k})$ have the following forms 
\begin{align}
\label{polarization2}
e^{+}_{ij}(\mathbf{k}) =\frac{1}{\sqrt{2}} \left( \begin{array}{ccc}
\sin^2{\theta} & -\sin{\theta}\cos{\theta} & 0 \\
-\sin{\theta}\cos{\theta} & \cos^2{\theta} & 0 \\
0 & 0 & -1 \\
\end{array} \right) ,
e^{\times}_{ij}(\mathbf{k}) = \frac{i}{\sqrt{2}}\left( \begin{array}{ccc}
0 & 0 & -\sin{\theta} \\
0 & 0 & \cos{\theta} \\
-\sin{\theta} & \cos{\theta} & 0 \\
\end{array} \right) ~.
\end{align}
Plugging the above polarization matrices in Eq. (\ref{tensor}),  the Fourier mode of the tensor field is given by
\begin{align}
\label{polarization3}
\widehat{h}_{ij}(\mathbf{k}) =\frac{1}{\sqrt{2}} \left( \begin{array}{ccc}
\widehat{h}_{+}\sin^2{\theta} & -\widehat{h}_{+}\sin{\theta}\cos{\theta} & -i\widehat{h}_{\times}\sin{\theta} \\
-\widehat{h}_{+}\sin{\theta}\cos{\theta} & \widehat{h}_{+}\cos^2{\theta} & i\widehat{h}_{\times}\cos{\theta} \\
-i\widehat{h}_{\times}\sin{\theta} &  i\widehat{h}_{\times}\cos{\theta}& -\widehat{h}_{+} \\
\end{array} \right) \, .
\end{align}
This expression will be used in the following  when  we calculate the cross-correlation between the tensor mode and the curvature perturbation as well as with the gauge field.

The tensor excitations has the standard profile 
\ba
\label{hs-wave}
{h}_{s}(k,\tau) = \frac{2 i H\tau }{M_{P}\sqrt{2 k}}\left(1-\frac{i}{ k\tau} \right)e^{-ik\tau}  \, ,
\ea
yielding the power spectrum 
\ba
\big \langle   \widehat{h}_{ij}(\mathbf{k_1})  \widehat{h}_{ij}(\mathbf{k_2})  \big \rangle
= (2 \pi)^3 \delta^{(3)} (\bfk_1 + \bfk_2) P_h(k_1) \, .
\ea

In the absence of anisotropy the fraction of tensor to scalar power spectrum  $r\equiv 2 P_h/P_\calR$
is given by $ r= 16 \epsilon$ with the observational bound \cite{Ade:2015lrj} $r < 0.1$.


\subsection{The mixing interactions containing $h_{ij}$}

\begin{figure}
  \hspace{0cm}
  \includegraphics[width=1\linewidth]{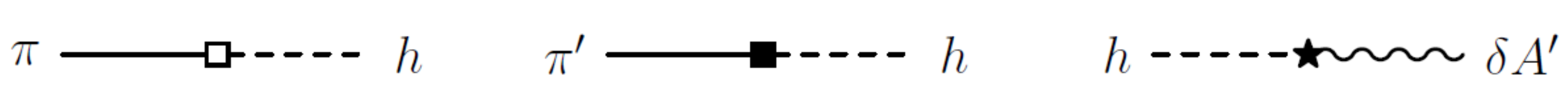}
  \caption{The exchange vertices involving the mixings of tensor-scalar and tensor-gauge field. The tensor field is denoted by the dashed line.   }
  \label{tensor-mix}
\end{figure}

Now we present the interactions mixing $\pi$ and $\delta A_i$ with $h_{ij}$.  These mixing are presented  schematically by Feynman diagram in Fig. \ref{tensor-mix}.  Note that the interaction 
between $\pi$ and $h_{ij}$ induces a non-zero $\langle \pi h_{ij} \rangle$
cross correlation. In addition, the interaction between $\delta A_i$ and $h_{ij}$ contributes  to the cross correlation $\langle \pi h_{ij} \rangle$ via exchanging  two $\delta A_i$ perturbations.  

Compared to the interactions involving scalar and gauge field perturbations studied in previous section, there is one difference now. In the previous sections, the scalar perturbations interacted with the scalar polarization of the gauge field fluctuations, $\delta A_{(S)}= (0,  \delta A_x, \delta A_y, 0)$ while it did not interact with the vector sector of the gauge field fluctuations, $\delta A_{(V)}= (0,  0, 0, \delta A_{V})$. However, as we shall see below, the tensor perturbations interact with both  $\delta A_{(S)}$ and $\delta A_{(V)}$ fluctuations.  Note that from the Coulomb radiation gauge,  we have $k_y \delta A_y = - k_x \delta A_x$ so we can choose either of $\delta A_x$ or $\delta A_y$, in addition with $\delta A_V$ perturbations. 

Now, collecting the contributions of tensor perturbations and its mixing with $\pi$ and $\delta A_i$ fluctuations to the first and second order perturbations of $X$ and $Z$ we have  
\ba
\delta X^{(1)}=4a^{-4}A^{\prime 2}_x h_{xx} , \quad \delta X^{(2)}=4a^{-4}\left( A^{\prime}_xh_{xx}\delta{A^{\prime}}_{x}+A^{\prime}_xh_{xy}\delta{A^{\prime}}_{y}+A^{\prime}_xh_{xz}\delta{A^{\prime}}_{V}\right) \, ,
\ea
and
\ba
\delta Z^{(1)}=-2a^{-4}A^{\prime 2}_x h_{xx} , \quad \delta Z^{(2)}=-2a^{-4}\left( A^{\prime}_xh_{xx}\delta{A^{\prime}}_{x}+A^{\prime}_xh_{xy}\delta{A^{\prime}}_{y}+A^{\prime}_xh_{xz}\delta{A^{\prime}}_{V}\right) \, .
\ea

Plugging these expressions into the EFT action (\ref{action-UG}) yields the following interactions involving various mixing between $\pi, h_{{ij}}$ and $\delta A_{i}$ fluctuations 
\ba
\label{inter-tensor}
S^{(2)}=\int d^{4} x \sqrt{-g}a^{-4}\left[\tilde M_{1}\left(A^{\prime}_xh_{xx}\delta{A^{\prime}}_{x}+A^{\prime}_xh_{xy}\delta{A^{\prime}}_{y}+A^{\prime}_xh_{xz}\delta{A^{\prime}}_{V}\right)+a^{-1}\tilde M_{1}^{\prime} A^{\prime 2}_x \pi h_{xx}\right] \, .
\ea

From the above quadratic action, the interaction Hamiltonians involving the mixing of tensor perturbations with the scalar and gauge field fluctuations 
have the following form  
\ba
\label{int Hamil}
H_{int} &=  H_{\pi h_+} + H_{\pi^{\prime} h_+}  + 
H_{\delta A_{x} h_{+}}+H_{\delta A_{V} h_\times}
\ea
in which 
\ba
\label{Hpi}
 H_{\pi h_+}  = 2 \sqrt{2}\,   \overline{M}_1\overline{A}^2H
\sin^2{\theta} a^4\pi {h}_{+} , \quad 
\label{Hpi2}
  H_{\pi^{\prime} h_+} = 2 \sqrt{2}\, \overline{A}^2\overline{M}_2\sin^2{\theta} a^3\pi^{\prime} {h}_{+}
\ea
and
\ba\label{haaa}
 H_{h_+  \delta A_x} = 
 -\frac{\overline{M}_1\, \overline A}{\sqrt{2}}  \delta A^{\prime}_{x} {h}_{+} , \quad 
H_{h_\times  \delta A_{V}} = \frac{i \overline{M}_{1}\,  \quad 
\overline{A}}{\sqrt{2}}\sin{\theta}\delta A_V h_{\times} \, .
\ea
Note that we used the  gauge constraint $k_{x } \delta A_{x} = -k_{y } \delta A_{y}$ to simplify $H_{h_+  \delta A_x} $ as given in  Eq. (\ref{haaa}).  

We comment that the interaction $ H_{\pi^{\prime} h_+}$, containing the factor $\overline M_{2}$, 
does not exist in the quadratic Lagrangian (\ref{inter-tensor}). However, this interaction originates upon construction the Hamiltonian from the Lagrangian  (\ref{inter-tensor}) and after taking into account the anisotropic corrections in conjugate momentum. This is similar to correction in second order Hamiltonian in scalar sector in Eq. (\ref{2-interaction}).  Consequently $ H_{\pi^{\prime} h_+}$ (along with $ H_{\pi h_+} $) is second order in $\overline A^{2}$. As such, we can neglect their contributions in anisotropy in tensor power spectrum.  However, we can not neglect  their contributions in  scalar-tensor cross correlations a priori. As we shall see explicitly below, the contributions of the leading interactions $ H_{h_+  \delta A_x} $ and $H_{h_\times  \delta A_{V}}$ in the 
scalar-tensor cross correlations is in the form of a nested integral. Consequently,  the contributions of
$ H_{h_+  \delta A_x} $ and $H_{h_\times  \delta A_{V}}$ in scalar-tensor cross correlations goes like
$\overline A^{2}$, i.e. the same order as the direct contributions from the sub-leading interactions $ H_{\pi^{\prime} h_+}$ and 
$ H_{\pi h_+} $.

Having calculated the various interactions mixing $\pi$ and $\delta A_{i}$ with $h_{{ij}}$, we  
calculate in turns the anisotropy induced in tensor power spectrum and the 
$\langle \pi h_{ij} \rangle$ cross correlation.


\subsection{Anisotropic tensor power spectrum}
\label{tensor-power}
Here we calculate the anisotropy induced in tensor power spectrum. There are two sources for anisotropy in tensor power spectrum. The first source is the contribution from  $H_{\pi h_+}$ and $H_{\pi^{\prime} h_+}$ while the second contribution comes  from $H_{h_+  \delta A_x}$ and $H_{h_\times  \delta A_{V}}$. However, looking at the amplitudes of these interactions, we see that 
 $H_{\pi h_+}$ and $H_{\pi^{\prime} h_+}$
are at the order $\overline A^2 \propto R$ while the interactions $H_{h_+  \delta A_x}$ and $H_{h_\times  \delta A_{V}}$ are at the oder $\overline A \propto \sqrt{R}$. Therefore, in the limit of small anisotropy $R \ll 1$, we can safely neglect the contribution of $H_{\pi h_+}$ and $H_{\pi^{\prime} h_+}$ in tensor power spectrum anisotropy. With this approximation in mind, we calculate the  anisotropic tensor power spectrum  $\delta \big{\langle} {h}_{+} {h}_{+}\big{\rangle}$ and $\delta \big{\langle} {h}_{\times} {h}_{\times}\big{\rangle}$. For the former we have 
\ba
\label{tensor plus}
\delta \big{\langle} {h}_{+} {h}_{+}\big{\rangle} &\simeq& - \int_{\tau_{0}}^{\tau_{e}} d\tau_{1}\int_{\tau_{0}}^{\tau_{1}} d\tau_{2}\bigg{[}H_{\delta A_{x} {h}_{+ }  } ,\big{[}H_{\delta A_{x} {h}_{+}}, {h}_{+} {h}_{+}\big{]}\bigg{]}  \nonumber\\
&=&
-\frac{\Big(c_s \left(c_v^2-3\right)-c_v^2+1\Big)^2}{ 2 M_P^4 c_v k^3} 
\overline M_1\, \overline{A}^2 N^2 \sin^2{\theta} \, .
\ea
Similarly, we also obtain the same result for $\delta \big{\langle} {h}_{\times} {h}_{\times}\big{\rangle}$. 
We see that the two polarization of tensor perturbations behave symmetrically for anisotropy in tensor power spectrum.  

The total anisotropy in tensor power spectrum,  after replacing the combination $\overline M_1\, \overline{A}^2$ in term of $R$, is given by
\ba
\delta \big{\langle} {h} {h}\big{\rangle}_{\mathrm{tot}} = -3 \Big(c_s \left(c_v^2-3\right)-c_v^2+1\Big)^2 \frac{ R H^2 }{  M_P^2 c_v k^3}  N^2 \sin^2{\theta} \, .
\ea
We see the non-trivial appearance of $c_s$ and $c_v$ in anisotropy of tensor power spectrum.

For the simple case of Maxwell setup with $c_{s}= c_{v}=1$, we have
\ba
\delta \big{\langle} {h} {h}\big{\rangle}_{\mathrm{tot}} = -\frac{24 H^{2 }R}{ M_P^2k^3} N^2 \sin^2{\theta} \, .
\ea
The fractional correction to tensor power spectrum in this case  is
\ba
\frac{ \delta \big{\langle} {h} {h}\big{\rangle}_{\mathrm{tot}} }{P_{h}^{(0)}}  \simeq 24 R N^{2 } = \frac{\epsilon g_{*}}{2} = 8 r g_{*} \,,
\ea
in which, to obtain the final result, the relation  $r= 16 \epsilon$ has been used to eliminate the slow-roll parameter in favor of  $r$, the ratio of tensor to scalar power spectra. Taking the observational bound $r \lesssim 0.1$ and $| g_* | \lesssim 10^{-2} $ the above ratio is less
than $10^{-2}$. The prospect for detection such  an small anisotropy in tensor power spectrum is not promising. 


\subsection{Scalar-tensor cross correlation}

Here we calculate the scalar-tensor cross correlation. Note that only the polarization $s=+$ contributes to the scalar-tensor cross correlation. This is because  $\pi$ couples only to $s=+$ polarization and not to  $s=\times$ polarization.

The analysis here is somewhat more involved than 
the analysis of tensor anisotropy in previous sub-section. The reason is that not only $H_{h_+  \delta A_x}$  but also  $H_{\pi h_+}$ and $H_{\pi^{\prime} h_+}$  contribute to 
$\langle \pi h_{ij} \rangle $ cross correlation. Though $H_{\pi h_+}$ and $H_{\pi^{\prime} h_+}$
are  at the order $R$ and $H_{h_+  \delta A_x}$ is at the order $\sqrt R$, but the former Hamiltonians appear  linearly in in-in integrals while the latter Hamiltonian appears quadratically in  nested integrals. More specifically,  we have 
\ba
\Big{\langle} \pi_{\mathbf{k}_{1}}(\tau_{e}) {h}_{+\mathbf{k}_{2}}(\tau_{e})\Big{\rangle} &=& i \int_{\tau_{0}}^{\tau_{e}} d\tau_{1}\Big{[} H_{\pi {h}_{+}} , \pi_{\mathbf{k}_{1}}{h}_{+\mathbf{k}_{2}}\Big{]}+ i \int_{\tau_{0}}^{\tau_{e}} d\tau_{1}\Big{[} H_{\pi^{\prime} {h}_{+}} , \pi_{\mathbf{k}_{1}}{h}_{+\mathbf{k}_{2}}\Big{]}
\nonumber\\
&& - \int_{\tau_{0}}^{\tau_{e}} d\tau_{1}\int_{\tau_{0}}^{\tau_{1}} d\tau_{2}\Big{[}H_{\pi \delta A_{x}} ,\big{[}H_{\delta A_{x} {h}_{+}}, \pi_{\mathbf{k}_{1}}{h}_{+\mathbf{k}_{2}}\big{]}\Big{]}\nonumber\\
&& - \int_{\tau_{0}}^{\tau_{e}} d\tau_{1}\int_{\tau_{0}}^{\tau_{1}} d\tau_{2}\Big{[}H_{\pi^{\prime} \delta A_{x}} ,\big{[}H_{\delta A_{x} {h}_{+}}, \pi_{\mathbf{k}_{1}}{h}_{+\mathbf{k}_{2}}\big{]}\Big{]} \nonumber\\
&& -\int_{\tau_{0}}^{\tau_{e}} d\tau_{1}\int_{\tau_{0}}^{\tau_{1}} d\tau_{2}\Big{[} H_{\delta A_{x} {h}_{+}},\big{[} H_{\pi \delta A_{x}}, \pi_{\mathbf{k}_{1}}{h}_{+\mathbf{k}_{2}}\big{]}\Big{]} \nonumber\\
&& -\int_{\tau_{0}}^{\tau_{e}} d\tau_{1}\int_{\tau_{0}}^{\tau_{1}} d\tau_{2}\Big{[} H_{\delta A_{x} {h}_{+}},\big{[} H_{\pi^{\prime} \delta A_{x}}, \pi_{\mathbf{k}_{1}}{h}_{+\mathbf{k}_{2}}\big{]}\Big{]} \nonumber\\
&\equiv& \Big{\langle} \pi_{\mathbf{k}_{1}}(\tau_{e}) {h}_{+\mathbf{k}_{2}}(\tau_{e})\Big{\rangle}_{1} + \Big{\langle} \pi_{\mathbf{k}_{1}}(\tau_{e}) {h}_{+\mathbf{k}_{2}}(\tau_{e})\Big{\rangle}_{2}\nonumber\\
&& + \Big{\langle} \pi_{\mathbf{k}_{1}}(\tau_{e}) {h}_{+\mathbf{k}_{2}}(\tau_{e})\Big{\rangle}_{3} + \Big{\langle} \pi_{\mathbf{k}_{1}}(\tau_{e}) {h}_{+\mathbf{k}_{2}}(\tau_{e})\Big{\rangle}_{4}\nonumber\\
&& + \Big{\langle} \pi_{\mathbf{k}_{1}}(\tau_{e}) {h}_{+\mathbf{k}_{2}}(\tau_{e})\Big{\rangle}_{5} + \Big{\langle} \pi_{\mathbf{k}_{1}}(\tau_{e}) {h}_{+\mathbf{k}_{2}}(\tau_{e})\Big{\rangle}_{6} \, .
\ea

Each integral, respectively is calculated to be 
\ba
\Big{\langle} \pi_{\mathbf{k}_{1}}(\tau_{e}) {h}_{+\mathbf{k}_{2}}(\tau_{e})\Big{\rangle} _1&=&
\frac{2 \sqrt 2 c_s \overline M_1\overline{A}^2}{M_P^2B_1 H k^3}
\sin^2{\theta}N \\
\Big{\langle} \pi_{\mathbf{k}_{1}}(\tau_{e}) {h}_{+\mathbf{k}_{2}}(\tau_{e})\Big{\rangle} _2&=&
\frac{- \sqrt 2 \left(c_s^3-3 c_s^2-3 c_s+1\right)}{3 M_P^2B_1 c_s H k^3} \overline{A}^2\overline{M}_2 \sin^2{\theta}N  \\
\Big{\langle} \pi_{\mathbf{k}_{1}}(\tau_{e}) {h}_{+\mathbf{k}_{2}}(\tau_{e})\Big{\rangle} _3&=&
\frac{-c_s^2 \left(c_s \left(c_v^2-3\right)-c_v^2+1\right)}{ \sqrt 2 M_P^2 B_1 c_v H k^3}
\left(\overline{A}^2 \, \overline{M}_1\right)\sin^2{\theta}
N^2 \\
\Big{\langle} \pi_{\mathbf{k}_{1}}(\tau_{e}) {h}_{+\mathbf{k}_{2}}(\tau_{e})\Big{\rangle} _4&=&
\frac{-3 c_s^2 \left(c_s \left(c_v^2-3\right)-c_v^2+1\right)}{\sqrt{2}M_P^2B_1 c_v  Hk^3}\overline A^2\overline{M}_2 \sin^2{\theta} N^2 \\
\Big{\langle} \pi_{\mathbf{k}_{1}}(\tau_{e}) {h}_{+\mathbf{k}_{2}}(\tau_{e})\Big{\rangle} _5 &=&
-\frac{c_s^2 \left(c_s \left(c_v^2-3\right)-c_v^2+1\right)}{\sqrt 2 M_P^2 B_1 c_v H k^3}
\overline{M}_1\overline{A}^2
\sin^2{\theta}N^2 \\
\Big{\langle} \pi_{\mathbf{k}_{1}}(\tau_{e}) {h}_{+\mathbf{k}_{2}}(\tau_{e})\Big{\rangle} _6&=&
\frac{9 c_s^4}{4 M_P^2B_1^2 c_v H k^3}\frac{1}{\sqrt{2}}\overline{A}^2\overline{M}_2 
\sin^2{\theta} N^2
\ea

Now adding the above six contributions, the total scalar-tensor cross correlation is given by 
\ba
&&\Big{\langle} \pi_{\mathbf{k}_{1}}(\tau_{e}) {h}_{+\mathbf{k}_{2}}(\tau_{e})\Big{\rangle} _{\mathrm{tot}}= \frac{\sqrt{2}\,  \overline A^2 N}{3M_P^2 B_1 H k^3 c_s} \bigg[-6 \overline M_1c_s^2-\overline M_2 \left(c_s+1\right) \left(\left(c_s-4\right) c_s+1\right)\bigg] \sin^2{\theta}\nonumber\\
&&~~~~~ -\frac{\overline A^2 c_s^2 N^2 }{4 M_P^2\sqrt{2} B_1^2 H k^3 c_v}\bigg[4 B_1 \left(2 \overline M_1+3\overline M_2\right) \left(\left(c_s-1\right) c_v^2-3 c_s+1\right)-9 \overline M_2 c_s^2\bigg] \sin^2\theta
\ea
Barring the accidental cancellation between $\overline M_1$ and $\overline M_2$, the second term is larger by an additional factor of $N \gg 1$ and one may neglect the contribution of the first term. 
As before, we obtain a non-trivial  appearance of  $c_s, c_v, \overline M_1$ and $\overline M_2$.   Finally, the scalar-tensor cross correlation has the quadrupole anisotropy with its amplitude  scaling like $R N^2$. 

As an example, for the case of Maxwell theory with $c_s=c_v=1$, we have 
\begin{align}
\Big{\langle} \pi_{\mathbf{k}_{1}}(\tau_{e}) {h}_{+\mathbf{k}_{2}}(\tau_{e})\Big{\rangle} _{\mathrm{tot}}
\simeq -\frac{12\sqrt{2}R}{\epsilon M_P^2 H k^3}N^2\sin^2{\theta} \, .
\label{Maxwell-pih}
\end{align}
It is constructive to compare the amplitude of scalar-tensor cross correlation Eq. (\ref{Maxwell-pih}) with the amplitude of scalar perturbations. Using the above result we have,
\ba
\frac{\Big{\langle} H \pi_{\mathbf{k}_{1}}(\tau_{e}) {h}_{+\mathbf{k}_{2}}(\tau_{e})\Big{\rangle} _{\mathrm{tot}}}{P_{\calR}^{(0)}} \simeq 96 \sqrt2 R N^2 =2 \sqrt 2 \epsilon |g_*|   = \frac{\sqrt 2 }{8} 
r |g_*| \, .
\ea
Note that we have multiplied $\pi$ by an additional factor $H$ since the curvature perturbations 
$\calR$ is related to $\pi$ by an additional factor $H$. 
 As discussed before,  taking the observational bound $r \lesssim 0.1$ and $| g_* | \lesssim 10^{-2} $ the above ratio is less than $10^{-3}$ .  This seems too small to have interesting observational effects. However, if one consider scenarios beyond simple Maxwell theory  with large enough amplitude of scalar-tensor cross correlation, interesting observational imprints can be obtained from the anisotropic structure of $TB$ and $EB$ cross correlations in CMB maps \cite{Watanabe:2010bu, Chen:2014eua}.

Before closing this section we comment that one can also look for anisotropy in scalar power spectrum $ \langle \pi^2 \rangle$ induced from the scalar-tensor mixing. On the physical ground, one expects these corrections come from the nested integrals with two  scalar-tensor interaction Hamiltonians via
\ba
\label{delta-P-zeta-b}
\delta \big \langle \pi_\bfk \pi_\bfk^*  \big\rangle  = - \int_{\tau_{0}}^{\tau_{e}} d\tau_{1} \int_{\tau_{0}}^{\tau_{1}}d\tau_{2} \left \langle
\Big{[} H_{I}(\tau_{2}) , \big{[} H_{I}(\tau_{1}) ,
\pi_\bfk (\tau_e) \pi_\bfk^*(\tau_e)  \big{]}\Big{]} \right \rangle  \, , 
\ea
where the interaction Hamiltonian are given in (\ref{Hpi}). However, one can see from the coefficients of these Hamiltonians that these corrections are at the order of $R^2$ and are much suppressed compared to anisotropy obtained in section \ref{power} which was at the order of $R$.


\section{Remarks on Geometric Approach to EFT }
\label{old approach}

As mentioned before, the first attempt to construct the EFT of anisotropic inflation involving the gauge field was performed in \cite{Abolhasani:2015cve}. The main insight in that approach was to choose the 
unitary gauge such that all matter perturbations from $\delta \phi$ and $\delta A^\mu$ are turned off and all perturbations are encoded in metric parts. This is in line with the original approach of EFT for single field inflation \cite{Cheung:2007st} .  In comparison, in our current analysis we have followed the approach of EFT in multiple field models as in  \cite{Senatore:2010wk} with the guidelines from the   Weinberg's approach \cite{Weinberg:2008hq} to read off the dominant interactions. As we have seen, this practical approach was efficient to construct the three-point interactions relevant to calculate the bispectrum. In addition, this approach allows one easily to construct the interactions beyond three point functions to calculate trispectrum and beyond. Here we comment on the approach used in  \cite{Abolhasani:2015cve} and difficulty associated with that approach to go beyond the two-point interactions.

In the approach of \cite{Abolhasani:2015cve} the unitary gauge is defined such that matter perturbations from $\delta \phi$ and $\delta A^\mu$ fluctuations are turned off and all perturbations are 
encoded in metric parts. Besides the invariance under the general coordinate transformation
\ba
\label{coordinate}
x^\mu \rightarrow x'^\mu=   x^\mu + \xi^\mu(x^\nu) \, ,
\ea
one has to implement the invariance of gauge field  under the $U(1)$
gauge transformation
\ba
\label{U1-gauge}
A^\mu \rightarrow A^\mu + \nabla^\mu  {\cal F} = A^{\mu} + g^{\mu \nu} \partial_\nu {\cal F} \, ,
\ea
in which  ${\cal F} (x^\nu)$ is a scalar.

Now let us  decompose the four vector $\xi^\mu$ into the transverse and the longitudinal parts,  
$\xi^\mu_T$ and $\xi^\mu_L$  as follows
\begin{equation}
\xi ^\mu = \nabla ^\mu \xi _L+\xi ^\mu _T=g^{\mu\alpha}\partial _\alpha \xi _L + \xi ^\mu _T \, ,
\end{equation}
subject to $\nabla _\mu \xi ^\mu _T=0$.   

Plugging these decompositions into gauge field transformation Eq. (\ref{U1-gauge}) yields 
\ba
\delta A^\mu  &\rightarrow  &
 \delta A^\mu +  A^x\partial _1 \xi ^\mu _T+g^{\mu\alpha}\partial _\alpha {\cal F} +  A^x\partial _1 \left(g^{\mu\alpha}\partial _\alpha \xi _L\right) , \nonumber \\
&=& \delta A ^\mu +  A^x\partial _1 \xi ^\mu _T+  A^x\left(\partial _1 g^{\mu \alpha}\right)\partial _\alpha \xi _L+
 \dot{  A}^x g^{\mu 0}\partial _1 \xi _L + g^{\mu \alpha} \partial _\alpha \left(  A^x\partial _1 \xi _L + {\cal F} \right).
\ea
Note that the above  transformation encodes both the $U(1)$ transformation (\ref{U1-gauge}) and the general coordinate transformation Eq. (\ref{coordinate}). 

As shown in \cite{Abolhasani:2015cve} imposing the unitary gauge $\delta \phi = \delta A^\mu=0$, the full 
four dimensional diffeomorphism and the gauge invariance is reduced to a smaller symmetry $x^\mu \rightarrow x^\mu +\xi ^\mu$ in which
\begin{align}
\xi ^\mu _T=\xi ^\mu _T(t,y,z), \qquad \left(\partial _1 g^{\mu \alpha}\right) \partial _\alpha \xi _L=\frac{\dot{ A^x }}{ A^x}g^{\mu 0}\partial _1 \xi _L,    \quad \quad \mathrm{ (remnant ~  symmetry) } \, 
\label{UG-symmetry}
\end{align}
excluding $\xi ^0$ component. 

Note that the above remnant symmetry is smaller than the remnant symmetry 
$x^i \rightarrow x^i + \xi( t, x, y, z)$ in unitary gauge in isotropic model of single scalar field \cite{Cheung:2007st}. This in turn induces restrictions in constructing interactions consistent with the unitary gauge in our setup containing gauge field and inflaton field.

The goal in the approach of \cite{Abolhasani:2015cve} is to construct the invariant operators from the metric perturbations and their derivatives.  Under the general coordinate transformation Eq. (\ref{coordinate}), the component of metric $g_{\alpha \beta}$ changes as
\ba
g_{\alpha\beta}(x)=\frac{\partial x^{\prime\gamma}}{\partial x^{\alpha}}\frac{\partial x^{\prime\sigma}}{\partial x^{\beta}}g^\prime_{\gamma\sigma}(x^\prime)=(\Lambda^{-1})^{\gamma}_{\alpha}(\Lambda^{-1})^{\sigma}_{\beta}\ g^\prime_{\gamma\sigma}(x^\prime),
\ea
in which $\Lambda^\alpha_\gamma \equiv \frac{\partial x^{\alpha}}{\partial{x^{\prime\gamma}}}$ 
and 
\ba
\label{Lambda-inverse}
(\Lambda^{-1})^{\alpha}_{\gamma}(x^\prime)=(\delta^{\alpha}_{\gamma}+\partial_{\gamma}\xi^{\alpha}+\partial_{\gamma}\xi^{\mu}\partial_{\mu}\xi^{\alpha}
+\partial_{\gamma}\xi^{\mu}\partial_{\mu}\xi^{\nu}\partial_{\nu}\xi^{\alpha}+...),
\ea
In particular, note that the partial derivatives in the second line are with respect to $x^\prime$. 

As in single field model, a good building block consistent with the remnant symmetry (\ref{UG-symmetry}) is $\delta g^{00}$. In addition, one may consider $\delta g_{1\alpha}$ as another building block. However, as shown in  \cite{Abolhasani:2015cve},  $\delta g_{1\alpha}$ is not a four vector under remnant symmetry. Therefore one should find a combination of $g_{1\alpha}$ and its derivative to construct an invariant scalar or a covariant  tensor to start with. This was constructed in \cite{Abolhasani:2015cve} in which it was shown that the quantity defined via
\begin{equation}
G_{\alpha \beta} \equiv \partial _\alpha  g_{\beta 1}-\partial _\beta g_{\alpha 1} +\frac{{\dot{A}^x}}{ {A^x} }\left(\delta ^0_{\alpha} g_{\beta 1}-\delta ^0_{\beta}g_{\alpha 1}\right) \, ,
\end{equation}
not only respects  the remnant symmetry \eqref{UG-symmetry} but also it is a four-tensor under the linear coordinate transformation  Eq. (\ref{coordinate}), i.e. 
\begin{equation}
G_{\alpha \beta} =\Lambda ^{\alpha ^\prime}_{\alpha}  \Lambda ^{\beta ^\prime}_\beta G_{\alpha ^\prime \beta ^\prime} \, .
\end{equation}
Consequently, one can write down the starting action in unitary gauge from the four scalars constructed  from  the contractions of $G_{\alpha \beta}$. As shown in \cite{Abolhasani:2015cve} this is as far as one can go with the metric perturbations and their derivatives. 

So far everything is good in the approach employed in \cite{Abolhasani:2015cve}. However, this approach becomes problematic if one goes beyond power spectrum and try to look at higher order interactions. To be more specific, it is shown in \cite{Abolhasani:2015cve} 
that by expanding $\Lambda^{-1}$  up to the first order in $\xi$, the quantity $G_{\alpha\beta}$ is a four tensor under general coordinate transformation. However, it is no longer a four tensor under higher order general coordinate transformation.  This is because the derivatives in the  higher order terms  in $\Lambda^{-1}$ are with respect to $x'$ as mentioned after Eq. (\ref{Lambda-inverse}). To see the significance of this effect, note that under a coordinate transformation we will have
\ba
g^\prime_{\alpha 1}(x^\prime)&=&\frac{\partial x^{\mu}}{\partial x^{\prime\alpha}}\frac{\partial x^{\nu}}{\partial x^{\prime 1}}g_{\mu\nu}(x)\nonumber\\
	&=&\Lambda_\alpha^\mu(x)\frac{\partial x^{\nu}}{\partial (x^1-\xi^1)}g_{\mu\nu}(x)\nonumber\\
	&=&\Lambda_\alpha^\mu(x)\bigg(\delta_1^\nu+\partial_1\xi^\nu+\partial_1\xi^\rho\partial_\rho\xi^\nu+...\bigg)g_{\mu\nu}(x)\nonumber\\
	&=&\Lambda_\alpha^\mu(x)\bigg(g_{\mu 1}(x)+g_{\mu\nu}(x)\partial_1\xi^\nu+g_{\mu\nu}(x)\partial_1\xi^\rho\partial_\rho\xi^\nu+...\bigg).
\ea
The appearance of the last two terms above inside the big bracket indicate that $g_{\alpha 1}$ is not a four-vector. The approach employed in  \cite{Abolhasani:2015cve} was specifically constructed to get rid of the second term which is linear in $\xi^\mu$. However, this approach was not extended  to cure the last non-linear term. This is because  in  \cite{Abolhasani:2015cve}  they were interested in power anisotropy in which the non-linear term plays no role. Now  for the bispectrum and higher order analysis,  to cancel the unwanted non-linear term, one has to modify the definition of $G_{\alpha\beta}$ such that it remains a four tensor to higher orders. 
Having said this, we stress that the approach employed in  \cite{Abolhasani:2015cve} is perfectly valid to linear order in perturbations so it can be used for anisotropy in power spectrum as  studied in \cite{Abolhasani:2015cve}.

In the current work, in order to get away with this difficulty, we have employed the EFT approach within the setup of multiple field scenarios. In this practical approach, one simply turns off the inflaton fluctuations by defining the unitary gauge as $\delta \phi=0$. Then, all allowed interactions for gauge field perturbations constructed from $\delta F_{\mu \nu} $  are presented along with building blocks constructed  from metric  such as $\delta g^{00}$. As we have seen this approach is particularly useful for bispectrum and interactions involving tensor perturbations.

\section{Summary and Discussions}

In this work we have studied EFT in the setup of anisotropic inflation. In these scenarios, in addition to inflaton field, there is a background $U(1)$ gauge field which induces a preferred direction during inflation and breaks the rotational invariance. We are interested in the limit where the background electric field energy density is small but a constant fraction of the total energy density. For this to be an attractor solution, there should be a non-trivial coupling between the inflaton field and the gauge field.  

We have presented the most general action in the decoupling limit 
allowed in the unitary gauge $\delta \phi =0 $ captured by the couplings $M_1, P_1, M_2, P_2$ and so on. The leading contribution in anisotropy power spectrum comes from the interactions sourced by  operators involving $M_1$ and $P_1$. The power anisotropy is in the shape of  quadrupole  with its amplitude scaling like $R N^2 c_s^5/c_v$ in which $c_s$  is  the sound speed of scalar perturbations 
and  $c_v$ is  the speed of gauge field fluctuations. It is interesting that our method can naturally incorporate  the scenarios with non-trivial  $c_s$ and $c_v$. As we have speculated, these scenarios may arise within the setup of string theory in which a brane with $U(1)$ gauge fields confined to its world volume  moves ultra-relativistically inside the string theory 
compactification as in DBI inflation.  

Our approach was particularly useful to calculate the bispectrum. The leading interactions in generating 
bispectrum come from $M_1, P_1, M_2$ and  $P_2$ operators. The anisotropic bispectrum scales 
like $R N^3  c_s^6/c_v^2 $ with an specific anisotropic shape. Our results agree with the previous results for bispectrum in models of anisotropic inflation within the simple setup of Maxwell where 
only $M_1$ is non-zero. Our method is easily applicable to calculate the trispectrum in the general setup of anisotropic inflation.

In addition, we have looked at anisotropy induced from the tensor sectors. Since the background is anisotropic, the usual decoupling of scalar and tensor perturbations in two point functions does not 
hold any longer. In particular, we will have scalar-tensor cross correlation.  In addition, we will have quadrupole anisotropy in tensor power spectrum induced from the exchange of two gauge field fluctuations. Both the scalar-tensor cross correlation and the anisotropy 
in tensor power spectrum scales like $R N^2$ with non-trivial dependence on $c_c, c_v$ and $\overline M_2/\overline M_1$.

While our starting action contains interactions involving parity violating operators, however to simplify the analysis we have assumed that there is no parity violations and these operators are 
turned off. Having said this, it will be interesting to study the effects of parity violating interactions within the setup of anisotropic inflation. In these cases, the two polarization of tensor perturbations behave quite differently and the tensor perturbations acquire handedness \cite{Barnaby:2011vw, Dimopoulos:2012av, Namba:2015gja, Bartolo:2015dga, Bartolo:2014hwa, Caprini:2014mja, Mukohyama:2014gba}. 
 
With the general shapes and amplitudes of anisotropies in scalar power spectrum, bispectrum and also the scalar-tensor cross correlation calculated, it is an interesting question to search for their imprints  in CMB maps \cite{Watanabe:2010bu, Chen:2014eua}. In particular, the scalar-tensor cross correlation induces non-trivial $TB$ and $EB$ cross correlations on CMB maps which do not exist in usual isotropic models. If these cross-correlations are large enough, one may find a new window in search for possible violations of  isotropy in primordial universe. \\

{\bf Acknowledgements:} We would like to thank A. A. Abolhasani, M. Akhshik and R. Emami for many useful discussions. We are grateful to S. Mukohyama and C. Lin for helpful discussions and for collaborations at the early stage of this work. H. F. would like to thank Yukawa Institute for Theoretical Physics (YITP) for hospitality during the progress of this work.



\begin{thebibliography}{aa}


\bibitem{Cheung:2007st} 
  C.~Cheung, P.~Creminelli, A.~L.~Fitzpatrick, J.~Kaplan and L.~Senatore,
  JHEP {\bf 0803}, 014 (2008)
  [arXiv:0709.0293 [hep-th]].

\bibitem{Cheung:2007sv} 
  C.~Cheung, A.~L.~Fitzpatrick, J.~Kaplan and L.~Senatore,
  JCAP {\bf 0802}, 021 (2008), 
  [arXiv:0709.0295 [hep-th]].

\bibitem{Manohar:1996cq} 
  A.~V.~Manohar,
  Lect.\ Notes Phys.\  {\bf 479}, 311 (1997)
  [hep-ph/9606222].


\bibitem{Burgess:2007pt} 
  C.~P.~Burgess,
  Ann.\ Rev.\ Nucl.\ Part.\ Sci.\  {\bf 57}, 329 (2007)
  [hep-th/0701053].
  


\bibitem{Ade:2013nlj}
  P.~A.~R.~Ade {\it et al.}  [Planck Collaboration],
  Astron.\ Astrophys.\  {\bf 571}, A23 (2014). 

\bibitem{Ade:2015hxq} 
  P.~A.~R.~Ade {\it et al.} [Planck Collaboration],
  Astron.\ Astrophys.\  {\bf 594}, A16 (2016) .

\bibitem{Kim:2013gka}
J.~Kim and E.~Komatsu,
Phys.\ Rev.\ D {\bf 88}, 101301 (2013)
[arXiv:1310.1605 [astro-ph.CO]].

\bibitem{Planck:2013jfk} 
  P.~A.~R.~Ade {\it et al.} [Planck Collaboration],
  Astron.\ Astrophys.\  {\bf 571}, A22 (2014)
  [arXiv:1303.5082 [astro-ph.CO]].

  
\bibitem{Ade:2015lrj} 
  P.~A.~R.~Ade {\it et al.} [Planck Collaboration],
  Astron.\ Astrophys.\  {\bf 594}, A20 (2016), 
  [arXiv:1502.02114 [astro-ph.CO]].


\bibitem{Abolhasani:2015cve} 
  A.~A.~Abolhasani, M.~Akhshik, R.~Emami and H.~Firouzjahi,
  JCAP {\bf 1603}, no. 03, 020 (2016), 
  [arXiv:1511.03218 [astro-ph.CO]].

\bibitem{Cannone:2015rra} 
  D.~Cannone, J.~O.~Gong and G.~Tasinato,
  JCAP {\bf 1508}, no. 08, 003 (2015)
  [arXiv:1505.05773 [hep-th]].

\bibitem{Hidaka:2014fra} 
  Y.~Hidaka, T.~Noumi and G.~Shiu,
  Phys.\ Rev.\ D {\bf 92}, no. 4, 045020 (2015)
  [arXiv:1412.5601 [hep-th]].


\bibitem{Lin:2015cqa} 
  C.~Lin and L.~Z.~Labun,
  JHEP {\bf 1603}, 128 (2016), 
  [arXiv:1501.07160 [hep-th]].


\bibitem{Bartolo:2015qvr} 
  N.~Bartolo, D.~Cannone, A.~Ricciardone and G.~Tasinato,
  JCAP {\bf 1603}, no. 03, 044 (2016), 
  [arXiv:1511.07414 [astro-ph.CO]].



\bibitem{Senatore:2010wk} 
  L.~Senatore and M.~Zaldarriaga,
  JHEP {\bf 1204}, 024 (2012)
  [arXiv:1009.2093 [hep-th]].


\bibitem{Weinberg:2008hq} 
  S.~Weinberg,
  Phys.\ Rev.\ D {\bf 77}, 123541 (2008), 
  [arXiv:0804.4291 [hep-th]].



\bibitem{Emami:2015qjl} 
  R.~Emami,
  arXiv:1511.01683 [astro-ph.CO].\\
J.~Soda,
Class.\ Quant.\ Grav.\  {\bf 29}, 083001 (2012)
[arXiv:1201.6434 [hep-th]].\\
A.~Maleknejad, M.~M.~Sheikh-Jabbari and J.~Soda,
Phys.\ Rept.\  {\bf 528}, 161 (2013)
[arXiv:1212.2921 [hep-th]].

		
\bibitem{Watanabe:2009ct}
  M.~a.~Watanabe, S.~Kanno and J.~Soda,
  Phys.\ Rev.\ Lett.\  {\bf 102}, 191302 (2009)
  [arXiv:0902.2833 [hep-th]].
  
 \bibitem{Watanabe:2010fh}
  M.~a.~Watanabe, S.~Kanno and J.~Soda,
  Prog.\ Theor.\ Phys.\  {\bf 123}, 1041 (2010)
  [arXiv:1003.0056 [astro-ph.CO]].

 \bibitem{Soda1} 
J.~Ohashi, J.~Soda and S.~Tsujikawa,
JCAP {\bf 1312}, 009 (2013)
[arXiv:1308.4488 [astro-ph.CO], arXiv:1308.4488].

J.~Ohashi, J.~Soda and S.~Tsujikawa,
Phys.\ Rev.\ D {\bf 88}, 103517 (2013)
[arXiv:1310.3053 [hep-th]].

J.~Ohashi, J.~Soda and S.~Tsujikawa,
Phys.\ Rev.\ D {\bf 87}, 083520 (2013)
[arXiv:1303.7340 [astro-ph.CO]].

S.~Kanno, J.~Soda, M.~-a.~Watanabe,
JCAP {\bf 1012}, 024 (2010).
[arXiv:1010.5307 [hep-th]].

K.~Murata, J.~Soda,
JCAP {\bf 1106}, 037 (2011).
[arXiv:1103.6164 [hep-th]].

S.~Yokoyama and J.~Soda,
JCAP {\bf 0808}, 005 (2008);

K.~Yamamoto, M.~-a.~Watanabe and J.~Soda,
Class.\ Quant.\ Grav.\  {\bf 29}, 145008 (2012)
[arXiv:1201.5309 [hep-th]].
  
  A.~Ito and J.~Soda,
  Phys.\ Rev.\ D {\bf 92}, no. 12, 123533 (2015), 
  [arXiv:1506.02450 [hep-th]].

  A.~Ito and J.~Soda,
  JCAP {\bf 1604}, no. 04, 035 (2016), 
  [arXiv:1603.00602 [hep-th]].



\bibitem{Emami1}
R.~Emami, H.~Firouzjahi, S.~M.~Sadegh Movahed, M.~Zarei,
JCAP {\bf 1102 } (2011)  005.
[arXiv:1010.5495 [astro-ph.CO]].

R.~Emami and H.~Firouzjahi,
JCAP {\bf 1201}, 022 (2012)
[arXiv:1111.1919 [astro-ph.CO]].


  S.~Baghram, M.~H.~Namjoo and H.~Firouzjahi,
  JCAP {\bf 1308}, 048 (2013)
  [arXiv:1303.4368 [astro-ph.CO]].



  R.~Emami and H.~Firouzjahi,
  JCAP {\bf 1510}, no. 10, 043 (2015)
  [arXiv:1506.00958 [astro-ph.CO]].



\bibitem{Emami:2013bk}
R.~Emami and H.~Firouzjahi,
JCAP {\bf 1310}, 041 (2013)
[arXiv:1301.1219 [hep-th]].

\bibitem{Abolhasani:2013zya}
A.~A.~Abolhasani, R.~Emami, J.~T.~Firouzjaee and H.~Firouzjahi,
JCAP {\bf 1308}, 016 (2013)
[arXiv:1302.6986 [astro-ph.CO]].


\bibitem{Abolhasani:2013bpa} 
  A.~A.~Abolhasani, R.~Emami and H.~Firouzjahi,
  JCAP {\bf 1405}, 016 (2014), 
  [arXiv:1311.0493 [hep-th]].

 \bibitem{Chen:2014eua}
 X.~Chen, R.~Emami, H.~Firouzjahi and Y.~Wang,
 arXiv:1404.4083 [astro-ph.CO].

\bibitem{Bartolo:2012sd}
  N.~Bartolo, S.~Matarrese, M.~Peloso and A.~Ricciardone,
  Phys.\ Rev.\ D {\bf 87}, 023504 (2013)
  [arXiv:1210.3257 [astro-ph.CO]].
  
\bibitem{Dulaney:2010sq} 
  T.~R.~Dulaney and M.~I.~Gresham,
  Phys.\ Rev.\ D {\bf 81}, 103532 (2010), 
  [arXiv:1001.2301 [astro-ph.CO]].

  
  \bibitem{Shiraishi:2013vja}
M.~Shiraishi, E.~Komatsu, M.~Peloso and N.~Barnaby,
JCAP {\bf 1305}, 002 (2013)
[arXiv:1302.3056 [astro-ph.CO]].
  
\bibitem{Shiraishi:2013oqa} 
  M.~Shiraishi, E.~Komatsu and M.~Peloso,
  JCAP {\bf 1404}, 027 (2014), 
  [arXiv:1312.5221 [astro-ph.CO]].

\bibitem{Barnaby:2012tk} 
  N.~Barnaby, R.~Namba and M.~Peloso,
  Phys.\ Rev.\ D {\bf 85}, 123523 (2012)
  [arXiv:1202.1469 [astro-ph.CO]].




\bibitem{various}

  K.~Dimopoulos, M.~Karciauskas, D.~H.~Lyth and Y.~Rodriguez,
  JCAP {\bf 0905}, 013 (2009)
  [arXiv:0809.1055 [astro-ph]].


  A.~E.~Gumrukcuoglu, B.~Himmetoglu, M.~Peloso,
  Phys.\ Rev.\  {\bf D81}, 063528 (2010).
  [arXiv:1001.4088 [astro-ph.CO]].

K.~Yamamoto,
Phys.\ Rev.\ D {\bf 85}, 123504 (2012)
[arXiv:1203.1071 [astro-ph.CO]].
  
  H.~Funakoshi and K.~Yamamoto,
  Class.\ Quant.\ Grav.\  {\bf 30}, 135002 (2013)
  [arXiv:1212.2615 [astro-ph.CO]].

  
T.~Fujita and S.~Yokoyama,
JCAP {\bf 1309}, 009 (2013)
[arXiv:1306.2992 [astro-ph.CO]].


S.~R.~Ramazanov and G.~Rubtsov,
Phys.\ Rev.\ D {\bf 89}, 043517 (2014)
[arXiv:1311.3272 [astro-ph.CO]].


  S.~Nurmi and M.~S.~Sloth,
  JCAP {\bf 1407}, 012 (2014)
  [arXiv:1312.4946 [astro-ph.CO]].


  R.~K.~Jain and M.~S.~Sloth,
  JCAP {\bf 1302}, 003 (2013)
  [arXiv:1210.3461 [astro-ph.CO]].


  F.~R.~Urban,
  Phys.\ Rev.\ D {\bf 88}, 063525 (2013)
  [arXiv:1307.5215 [astro-ph.CO]].


M.~Thorsrud, D.~F.~Mota and S.~Hervik,
JHEP {\bf 1210}, 066 (2012)
[arXiv:1205.6261 [hep-th]].

S.~Bhowmick and S.~Mukherji,
Mod.\ Phys.\ Lett.\ A {\bf 27}, 1250009 (2012)
[arXiv:1105.4455 [hep-th]].

  S.~Hervik, D.~F.~Mota and M.~Thorsrud,
  JHEP {\bf 1111}, 146 (2011)
  [arXiv:1109.3456 [gr-qc]].



C.~G.~Boehmer, D.~F.~Mota,
Phys.\ Lett.\  {\bf B663}, 168-171 (2008).
[arXiv:0710.2003 [astro-ph]].

T.~S.~Koivisto, D.~F.~Mota,
JCAP {\bf 0808}, 021 (2008).
[arXiv:0805.4229 [astro-ph]].


D.~H.~Lyth and M.~Karciauskas,
JCAP {\bf 1305}, 011 (2013)
[arXiv:1302.7304 [astro-ph.CO]].

Tuan Q. Do and W. F. Kao,
Phys. Rev. D 84, 123009.\\
Tuan Q. Do, W. F. Kao, and Ing-Chen Lin,
Phys. Rev. D 83, 123002.

\bibitem{Shiraishi:2016wec} 
  M.~Shiraishi, N.~S.~Sugiyama and T.~Okumura,
  arXiv:1612.02645 [astro-ph.CO].


  M.~Shiraishi, J.~B.~Mu�oz, M.~Kamionkowski and A.~Raccanelli,
  Phys.\ Rev.\ D {\bf 93}, no. 10, 103506 (2016), 
  [arXiv:1603.01206 [astro-ph.CO]].

  M.~Shiraishi, N.~Bartolo and M.~Liguori,
  JCAP {\bf 1610}, no. 10, 015 (2016), 
  [arXiv:1607.01363 [astro-ph.CO]].

  M.~Shiraishi,
  Phys.\ Rev.\ D {\bf 94}, no. 8, 083503 (2016), 
  [arXiv:1608.00368 [astro-ph.CO]].




\bibitem{Naruko:2014bxa} 
A.~Naruko, E.~Komatsu and M.~Yamaguchi,
  JCAP {\bf 1504}, no. 04, 045 (2015)
  [arXiv:1411.5489 [astro-ph.CO]].


\bibitem{various2} 

  N.~Bartolo, S.~Matarrese, M.~Peloso and A.~Ricciardone,
  JCAP {\bf 1308}, 022 (2013)
  [arXiv:1306.4160 [astro-ph.CO]].

  N.~Bartolo, M.~Peloso, A.~Ricciardone and C.~Unal,
  JCAP {\bf 1411}, no. 11, 009 (2014)
  [arXiv:1407.8053 [astro-ph.CO]].

  M.~Akhshik, R.~Emami, H.~Firouzjahi and Y.~Wang,
  JCAP {\bf 1409}, 012 (2014)
  [arXiv:1405.4179 [astro-ph.CO]].

  M.~Akhshik,
  JCAP {\bf 1505}, no. 05, 043 (2015)
  [arXiv:1409.3004 [astro-ph.CO]].

  X.~Chen, R.~Emami, H.~Firouzjahi and Y.~Wang,
  JCAP {\bf 1504}, no. 04, 021 (2015)
  [arXiv:1408.2096 [astro-ph.CO]].

  A.~Ricciardone and G.~Tasinato,
  arXiv:1611.04516 [astro-ph.CO].


  X.~Li, S.~Wang and Z.~Chang,
  Eur.\ Phys.\ J.\ C {\bf 75}, no. 6, 260 (2015)
  [arXiv:1502.02256 [gr-qc]].

  C.~Pitrou, T.~S.~Pereira and J.~P.~Uzan,
  JCAP {\bf 0804}, 004 (2008)
  [arXiv:0801.3596 [astro-ph]].


  G.~Esposito-Farese, C.~Pitrou and J.~P.~Uzan,
  Phys.\ Rev.\ D {\bf 81}, 063519 (2010)
  [arXiv:0912.0481 [gr-qc]].
  


\bibitem{Emami:2016ldl} 
  R.~Emami, S.~Mukohyama, R.~Namba and Y.~l.~Zhang,
  arXiv:1612.09581 [hep-th].


\bibitem{Mukohyama:2016npi} 
  S.~Mukohyama,
  Phys.\ Rev.\ D {\bf 94}, no. 12, 121302 (2016)
  [arXiv:1607.07041 [hep-th]].

\bibitem{Fleury:2014qfa} 
  P.~Fleury, J.~P.~Beltran Almeida, C.~Pitrou and J.~P.~Uzan,
  JCAP {\bf 1411}, no. 11, 043 (2014)
  [arXiv:1406.6254 [hep-th]].



\bibitem{Talebian-Ashkezari:2016llx} 
  A.~Talebian-Ashkezari, N.~Ahmadi and A.~A.~Abolhasani,
  arXiv:1609.05893 [gr-qc].


\bibitem{Ackerman:2007nb} 
  L.~Ackerman, S.~M.~Carroll and M.~B.~Wise,
  Phys.\ Rev.\ D {\bf 75}, 083502 (2007)
  [Phys.\ Rev.\ D {\bf 80}, 069901 (2009)]
  [astro-ph/0701357].

\bibitem{Pullen:2007tu} 
  A.~R.~Pullen and M.~Kamionkowski,
  Phys.\ Rev.\ D {\bf 76}, 103529 (2007)
  [arXiv:0709.1144 [astro-ph]].

\bibitem{Maldacena:2002vr} 
  J.~M.~Maldacena,
  JHEP {\bf 0305}, 013 (2003)
  doi:10.1088/1126-6708/2003/05/013
  [astro-ph/0210603].


\bibitem{ArmendarizPicon:1999rj} 
  C.~Armendariz-Picon, T.~Damour and V.~F.~Mukhanov,
  Phys.\ Lett.\ B {\bf 458}, 209 (1999)
  [hep-th/9904075].

\bibitem{Garriga:1999vw} 
  J.~Garriga and V.~F.~Mukhanov,
  Phys.\ Lett.\ B {\bf 458}, 219 (1999)
  [hep-th/9904176].

\bibitem{Alishahiha:2004eh} 
  M.~Alishahiha, E.~Silverstein and D.~Tong,
  Phys.\ Rev.\ D {\bf 70}, 123505 (2004)
  [hep-th/0404084].

\bibitem{Dimopoulos:2011pe} 
  K.~Dimopoulos, D.~Wills and I.~Zavala,
  Nucl.\ Phys.\ B {\bf 868}, 120 (2013)
  [arXiv:1108.4424 [hep-th]].


\bibitem{Weinberg:2005vy}
  S.~Weinberg,
  Phys.\ Rev.\ D {\bf 72}, 043514 (2005)
  [hep-th/0506236].
  
\bibitem{Emami:2014tpa} 
  R.~Emami, H.~Firouzjahi and M.~Zarei,
  Phys.\ Rev.\ D {\bf 90}, no. 2, 023504 (2014),
  [arXiv:1401.4406 [hep-th]].

\bibitem{Watanabe:2010bu}
M.~-a.~Watanabe, S.~Kanno and J.~Soda,
Mon.\ Not.\ Roy.\ Astron.\ Soc.\  {\bf 412}, L83 (2011)
[arXiv:1011.3604 [astro-ph.CO]].

\bibitem{Barnaby:2011vw} 
  N.~Barnaby, R.~Namba and M.~Peloso,
  JCAP {\bf 1104}, 009 (2011), 
  [arXiv:1102.4333 [astro-ph.CO]].

\bibitem{Dimopoulos:2012av} 
  K.~Dimopoulos and M.~Karciauskas,
  JHEP {\bf 1206}, 040 (2012)
  [arXiv:1203.0230 [hep-ph]].
  
\bibitem{Namba:2015gja} 
  R.~Namba, M.~Peloso, M.~Shiraishi, L.~Sorbo and C.~Unal,
  arXiv:1509.07521 [astro-ph.CO].

\bibitem{Bartolo:2015dga} 
  N.~Bartolo, S.~Matarrese, M.~Peloso and M.~Shiraishi,
  JCAP {\bf 1507}, no. 07, 039 (2015)
  [arXiv:1505.02193 [astro-ph.CO]].

\bibitem{Bartolo:2014hwa} 
  N.~Bartolo, S.~Matarrese, M.~Peloso and M.~Shiraishi,
  JCAP {\bf 1501}, no. 01, 027 (2015)
  [arXiv:1411.2521 [astro-ph.CO]].

\bibitem{Caprini:2014mja} 
  C.~Caprini and L.~Sorbo,
  JCAP {\bf 1410}, no. 10, 056 (2014)
  [arXiv:1407.2809 [astro-ph.CO]].

\bibitem{Mukohyama:2014gba}
  S.~Mukohyama, R.~Namba, M.~Peloso and G.~Shiu,
  JCAP {\bf 1408} (2014) 036, 
  [arXiv:1405.0346 [astro-ph.CO]].


\end{thebibliography}
\end{document}